%% file: randomtemplates_paper.tex
\documentclass[aps,prd,twocolumn,groupedaddress,nofootinbib]{revtex4}

\pdfoutput=1
\usepackage {amsmath}
\usepackage {amssymb}
\usepackage {mathrsfs}
\usepackage {natbib}
\usepackage {latexsym}
\usepackage {graphicx}
\usepackage {dsfont}
\usepackage {rotating}
\usepackage {wasysym}
\usepackage {multirow}
\usepackage {hhline}
\usepackage{graphicx}
\usepackage{hyperref}
\usepackage{color}

\input{defs.tex}
\newcommand{\dcc}{LIGO-P080090-01-Z}

\input{tag.tex}

\begin{document}


\title{Random template banks and relaxed lattice coverings}

\author{C.~Messenger}
\email{chris.messenger@aei.mpg.de}
\author{R.~Prix}
\author{M.~A.~Papa}
\affiliation{Albert-Einstein-Institut}


\date{\commitDATE\\\mbox{\small \commitID}\\\mbox{\dcc}}

\begin{abstract}
  Template-based searches for gravitational waves are often limited by the computational cost associated with searching
  large parameter spaces.
  The study of \emph{efficient} template banks, in the sense of using the smallest number of templates, is therefore of great practical interest.
  The ``traditional'' approach to template-bank construction requires every point in parameter space to be \emph{covered} by at
  least one template, which rapidly becomes inefficient at higher dimensions.
  Here we study an alternative approach, where any point in parameter space is covered only with a given probability $\conf < 1$.
  We find that by giving up complete coverage in this way, large reductions in the number of templates are possible, especially at
  higher dimensions.
  The prime examples studied here are ``random template banks'', in which templates are placed randomly with uniform probability over
  the parameter space. In addition to its obvious simplicity, this method turns out to be surprisingly efficient.
  We analyze the statistical properties of such random template banks, and compare their efficiency to traditional lattice coverings.
  We further study ``relaxed'' lattice coverings (using $\Zn$ and $\Ans$ lattices), which similarly cover any signal
  location only with probability $\conf$.
  The relaxed $\Ans$ lattice is found to yield the most efficient template banks at low dimensions ($n\lesssim 10$),
  while random template banks increasingly outperform any other method at higher dimensions.
\end{abstract}

\pacs{}

\maketitle

%
\section{Introduction}
%
Matched filtering is the optimal linear detection method for signals in additive stationary noise. This method is widely used in
searches for gravitational waves (GWs) in the data of
ground based detectors (LIGO~\cite{2004NIMPA.517..154A,1999PhT....52j..44B,1992Sci...256..325A},
GEO~\cite{2002CQGra..19.1377W,2002CQGra..19.1835G}, VIRGO~\cite{2006CQGra..23S.635A}, TAMA~\cite{0264-9381-21-5-004})
as well as being among those proposed for future space based detectors ~\cite{LISA}.
The majority of signals that are being searched for
have predictable waveforms but unknown waveform parameters. Matched filtering
consists of processing the data with multiple waveforms (templates) each corresponding to a different set of waveform
parameters. The different templates are spaced based on a metric
on the space of signal parameters.  The metric defines a distance measure directly related to the loss in the matched filter signal-to-noise ratio
for a given template and signal~\cite{bala96:_gravit_binaries_metric,owen96:_search_templates,1999PhRvD..60b2002O,brady98:_search_ligo_periodic}.
Using the metric, a template bank, usually in the form of a lattice, can be placed on the space such that the loss between any
putative signal and at least a single template in the bank is less than a predefined maximum value.

Template placement for gravitational wave data analysis has proven to be a complicated and involved procedure even in
the relatively low--dimensional spaces already searched~\cite{2007PhRvD..76h2001A,abbott-2008,abbott:102004,abbott:062002,abbott-2007,abbott-grb070201}.
Here we discuss the possibility of adopting a seemingly far less complicated template placement method whereby we
\emph{randomly} position templates within our search space rather than placing them on a lattice.

It has recently been shown \cite{2007CQGra..24..481P} that constructing optimal template banks can be interpreted as an
instance of the mathematical \emph{sphere covering problem}, and that results from this field of research can usefully
be applied to template banks.
For instance, the hyper-cubic $\Zn$ lattice covering is known to become extremely inefficient at higher dimensions compared to
other lattices, in particular the $\Ans$ lattice, which provides a highly efficient covering for dimensions up to $n < 24$.

In practice, however, constructing lattice-based template banks often turns out to be problematic, due to the difficulties
associated with adapting lattice coverings to curved parameter spaces and performing coordinate transformations to avoid
non-constant metric components.
Furthermore, even the best lattice covering becomes increasingly inefficient at higher dimensions ($n \gtrsim 10$, say),
which makes this approach increasingly unsuitable for problems involving high-dimensional parameter-spaces.

A radically different approach to template-bank construction consists in relaxing the strict requirement of \emph{complete} coverage
for a given mismatch, and instead require coverage only with a certain confidence $\conf < 1$. This is a natural step for searches
that employ statistical detection techniques which always involve a finite false-dismissal probability.
Connected to this idea are new types of template banks, commonly referred to as ``stochastic'', which have recently been studied and
applied by various groups \cite{babak-2008,harry-2008b,harry-2008,vdB2008:_stochastic}.
Stochastic template banks are constructed by randomly placing templates on the parameter space, accompanied by a
``pruning'' step in which ``superfluous'' templates, which are deemed to lie too close to each other, are removed.

Here we study an even simpler approach, which we refer to as ``random template banks'', in order to distinguish it from stochastic banks.
This method consists of placing \emph{the right number of templates} randomly, with probability density dependent on the metric
determinant and without any additional pruning steps.
Apart from the practical advantage of relative simplicity, this allows one to analyze the properties of such random template banks
analytically and in great detail. For example, we can explicitly determine the number of templates $\Nrand$ required to achieve
any desired level of coverage confidence~$\conf$.
This paper presents the first detailed study of the properties of such template banks, and an explicit comparison of their
efficiency to traditional full-coverage lattice template banks.

Despite their simplicity, random template banks are found to achieve astonishing levels of efficiency compared to
traditional template banks, especially at higher dimensions. They outperform even the highly efficient $\Ans$ lattice in
dimensions above $n \sim 6 - 7$ for covering confidences $\conf$ in the range of $90\%-95\%$.

As a by-product of this study, we also analyze the properties of ``relaxed lattice'' coverings, which share a fundamental
feature with random template banks: for any signal location, the nominal covering mismatch is guaranteed only with probability
$\conf < 1$. This results in
a coarser lattice and therefore a reduction of the number of templates. We find that these relaxed lattices generally
result in the most efficient template banks at dimensions up to $n \sim 11$, where random template banks start to dominate.

The plan of this paper is as follows: first we review the general template-bank problem and traditional lattice-based template
banks in Sec.~\ref{sec:template-placement}.
In Sec.~\ref{sec:random-template-placement} we present a detailed analysis of random template banks: We
calculate their template densities, and compare them to traditional lattice coverings, and we investigate some of the
relevant statistical properties of random template banks.
In Sec.~\ref{sec:relaxed-lattices} we describe a modification to traditional lattice template banks, termed relaxed lattice covering.
Sec.~\ref{sec:discussion} provides a summary and discussion of the results.
%
\section{``Traditional'' template-bank construction}  
\label{sec:template-placement}
%
In this section we briefly review some fundamental concepts used in constructing template banks, namely the parameter-space metric
\cite{bala96:_gravit_binaries_metric,owen96:_search_templates,prix06:_searc}
and lattice coverings \cite{1999PhRvD..60b2002O,brady98:_search_ligo_periodic,2007CQGra..24..481P}.
One key feature of traditional template banks is that they require \emph{complete} coverage of the parameter space, i.e.\ no
point in parameter space is allowed to be further away from its closest template than a given maximal mismatch.

Consider an $n$-dimensional parameter space $\Sn$, with coordinates $\{\lambda^i\}_{i=1}^n$.
Each point $\lambda$ describes a set of parameters of a signal model, which we assume to be an accurate description of the
true signal family $s(t; \lambda)$.  Assume we measured data $x(t)$ containing a signal $s(t;\signal)$ in addition to Gaussian
additive noise $n(t)$, i.e.\ $x(t) = n(t) + s(t;\signal)$.
Typically one constructs a detection statistic of the data, $X(\lambda;\,x)$, say, which is a scalar representing the probability
that a signal with parameters $\lambda$ is present in the data $x(t)$. Due to the random noise fluctuations n(t), $X$ is a random
variable, but with the property that its expectation value $\bar{X}(\lambda; \signal) \equiv E[X(\lambda; x)]$ has a
maximum at the true location of the signal $\lambda = \signal$.
We can define a notion of \emph{mismatch}, or squared length of a parameter offset $\Delta \lambda = \lambda - \signal$,
as the relative loss in the expected detection statistic due to this offset, i.e.\
\begin{equation}\label{eq:mismatch}
  \mis(\Delta\lambda;\signal) = 1 - \frac{\bar{X}(\lambda;\signal)}{\bar{X}(\signal; \signal)}
  = g_{ij}(\signal)\,\Delta\lambda^i\Delta\lambda^j + \ldots \,,
\end{equation}
where we use automatic summation over repeated indices $i, j$ and the metric tensor $g_{ij}$ is defined via Taylor
expansion of the mismatch $m$ in the small offset $\Delta\lambda$.
Using this definition of the metric, the proper volume of the parameter space $\Sn$ can now be expressed as
\begin{equation}\label{eq:proper_volume}
  \Vspace = \int_{\Sn} \, d V \,,\quad\mbox{with}\quad d V \equiv \sqrt{g}\; d^n\!\lambda\,.
\end{equation}
where $g\equiv\det g_{ij}$ is the determinant of the metric $g_{ij}$.
A common approach to the problem of parameter-space covering is to use a lattice of templates. Template-based
searches are often computationally expensive due to the large number of templates required, therefore much effort has gone into identifying
the most efficient covering, namely the lattice that requires the fewest templates to achieve complete coverage of the
parameter space~\cite{2007CQGra..24..481P}.
Note that constructing lattice template banks in curved parameter spaces is highly impractical, and most of the following results
implicitly assume that the parameter-space metric $g_{ij}$ is \emph{flat}, i.e.\ we can find coordinates in
which $g_{ij}$ is constant (i.e.\ independent of parameter-space location $\lambda$).

A parameter-space point $\lambda$ is considered to be ``covered'' by a template $\lambda_{(k)}$ if its squared distance to the template
is smaller than the given nominal mismatch $\misnom$, i.e.\
\begin{equation}
  \label{eq:1}
  g_{ij} \, \Delta\lambda_{(k)}^i \, \Delta\lambda_{(k)}^j < \misnom\,,\quad\mbox{with}\quad \Delta\lambda_{(k)}\equiv\lambda - \lambda_{(k)}\,.
\end{equation}
This is equivalent to saying that $\lambda$ lies within the $n$-dimensional sphere of radius $R = \sqrt{\misnom}$ centered on the
template $\lambda_{(k)}$. The construction of efficient (complete) template banks is therefore an instance of the \emph{sphere covering problem},
which asks for the sphere arrangement requiring the smallest number of overlapping spheres to completely cover an $n$-dimensional (Euclidean)
space~\cite{2007CQGra..24..481P,CONWAYSLOANE}.

A key quantity used in assessing the efficiency of a given sphere covering is its \emph{thickness}. 
The thickness $\Thickness$ is defined \cite{CONWAYSLOANE} as \emph{the average number of $n$-dimensional spheres (templates)
  covering any point in the parameter space}. For a complete coverage, the thickness therefore satisfies by definition
$\Thickness \ge 1$ (where in practice equality can only be attained for $n=1$, in higher dimensions there will always be some
overlap between spheres).
For a lattice covering, the thickness can be conveniently expressed as
\begin{equation}\label{eq:theta_lattice}
  \Thickness = \frac{\Vn \, \misnom^{n/2}}{\Vlattice(\misnom)}\,,
\end{equation}
where $\Vn$ is the volume enclosed by an $n$-dimensional unit-sphere \footnote{We are using the geometers convention
with respect to the definition of the $n$-dimensional sphere (or n-sphere) where the 1-sphere represents two points on a line, 
the 2-sphere is a circle, etc.}, namely
\begin{equation}
  \Vn = \frac{\pi^{n/2}}{\Gamma(n/2+1)},
\end{equation}
and $\Vlattice$ is the volume of a fundamental region of the lattice $\Lattice$, with covering radius $R = \sqrt{\misnom}$.
Note that under a linear rescaling $c$, lengths change like $R' = c \, R$, mismatches like $m' = c^2\,m$, and
lattice volumes like $\Vlattice' = c^n\, \Vlattice$. Therefore we see from Eq.~\eqref{eq:theta_lattice} that the thickness $\Thickness$ is a
scale-invariant property, characterizing the geometric \emph{structure} of a covering. In particular $\Thickness$ is independent
of mismatch $\misnom$.
A special instance of a fundamental lattice
region is the \emph{Voronoi cell} (also known as the \emph{Wigner-Seitz cell}), which is the set of points that are closer
to a given template than to any other template.  Let us also define at this point the volume covered by a single template as
\begin{equation}
  \label{eq:15}
  \Vtemplate \equiv \Vn \,\misnom^{n/2}\,.
\end{equation}
In the following it will also be useful to introduce the \emph{normalized thickness},
\begin{equation}
  \label{eq:2}
  \thickness \equiv \frac{\Thickness}{\Vn}\,,
\end{equation}
which corresponds to the number of templates per unit volume in the case of $\misnom = 1$. Like the thickness $\Thickness$,
this is a scale-invariant property of a covering, independent of mismatch $\misnom$.
As shown in \cite{2007CQGra..24..481P}, the total number of templates $N$ of a covering can be expressed in terms of the
normalized thickness as
\begin{equation}\label{eq:N_lattice_templates}
  N = \thickness\, \misnom^{-n/2} \, \Vspace\,,
\end{equation}
which shows that the total number of templates is proportional to the normalized thickness.

In the following we will focus on two lattices, namely the $\Zn$ (hyper-cubic) and the $\Ans$ lattice, known
respectively for their simplicity and covering efficiency. The normalized thickness is known analytically for both lattices,
namely
\begin{align}
  \thickness_{\Zn} &=\frac{n^{n/2}}{2^{n}}, \label{eq:cubic_thickness}\\
  \thickness_{\Ans}&=\sqrt{n+1}\left[\frac{n(n+2)}{12(n+1)}\right]^{n/2}.\label{eq:An_thickness}
\end{align}
In the following we will mostly use the normalized thickness for comparing different covering strategies, since it is proportional
to the total number of templates (Eq.~(\ref{eq:N_lattice_templates})), which is the quantity we wish to minimize in order to reduce the
computational cost of searching a parameter space.
%
\section{Random template banks}
\label{sec:random-template-placement} 
%

We now investigate the properties of a new type of template bank, which we call the ``random template bank'', which consists
of $\Nrand$ templates placed randomly with uniform probability distribution (per proper volume) over the parameter space $\Sn$.
Note that contrary to the lattice template-banks discussed in the previous section, nothing in the following requires the
parameter-space metric $g_{ij}(\lambda)$ to be constant or flat. The only practical assumption we will make for simplicity is that the
metric curvature radius has to be large compared to the covering radius of one template, so we can neglect metric curvature in the expression
for the volume of a single template given by Eq.~(\ref{eq:15}). Some practical issues arising from non-constant metrics will be discussed in
Sec.~\ref{sec:nonflat-spaces}.

\subsection{Number of required random templates $\Nrand$}
\label{sec:numb-requ-rand}

\begin{figure}
  \begin{center}
    \includegraphics[width=\columnwidth]{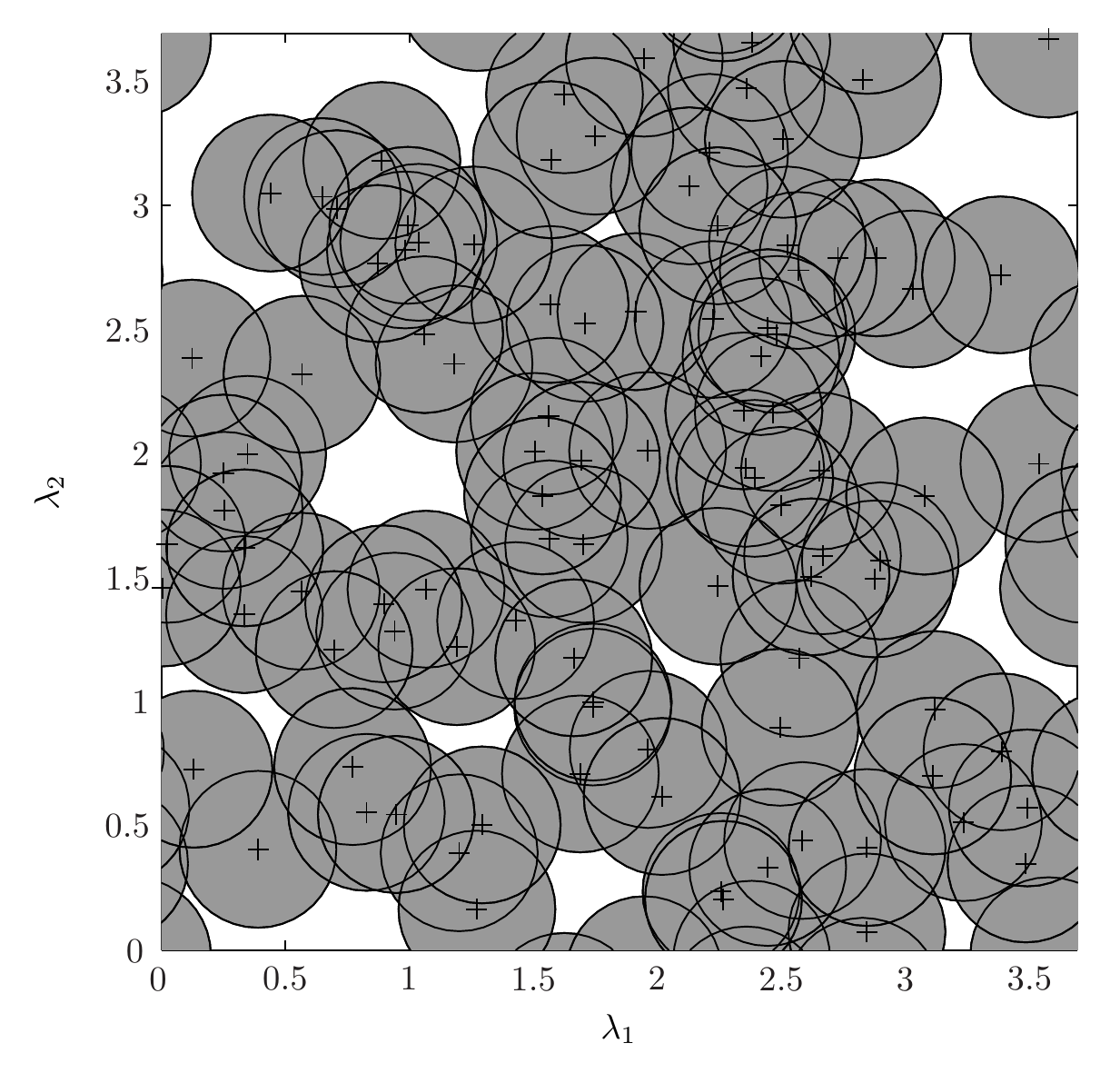}
    \caption{Example realization of a random template bank $^{0.9}\rand_2(0.1)$ in $n=2$ dimensions (using periodic boundary
      conditions), with nominal mismatch $\misnom=0.1$ and covering confidence $\conf=0.9$.
      The parameter-space volume $\Vspace$ was chosen such that the resulting number of templates is $\Nrand=100$.
      Template locations are indicated by crosses and template boundaries (corresponding to the covering radius
      $R=\sqrt{\misnom}$) are shown as black circles.  The covered and uncovered volumes are shaded grey and white
      respectively.
    }
    \label{fig:example_2d_covering}
  \end{center}
\end{figure}

Let us select a point $\signal \in \Sn$ which we assume to be the location of a signal. We assume that the covering sphere
with radius $R = \sqrt{\misnom}$ centered on $\signal$ does not intersect the boundary of the parameter-space $\Sn$, which allows
us to neglect boundary effects in the following discussion.
Now consider a single randomly placed template with uniform probability distribution (per proper volume).
What is the probability that this template does \emph{not} cover (``miss'') within $\misnom$ the
point $\signal$? This is equivalent to the probability that the template does not fall within the covering sphere centered on
$\signal$. The probability of falling within this volume is $\Vtemplate/\Vspace$, and so the answer is simply
\begin{equation}\label{eq:P0_of_mu}
  P\left(\miss(\misnom)\, |\,\Sn,\,\Nrand=1\right) = 1 - \frac{\Vn \,\misnom^{n/2}}{\Vspace}\,.
\end{equation}
If we were to place $\Nrand$ templates randomly in this way, the probability that \emph{none} of the templates
cover the point $\signal$ is therefore
\begin{equation}\label{eq:P0Nr_of_mu}
  P\left(\miss(\misnom)\, |\Sn,\, \Nrand\right) = \left(1 - \frac{\Vn \,\misnom^{n/2}}{\Vspace}\right)^{\Nrand}\,,
\end{equation}
since in this construction each template location is independent of all previously placed templates.  It follows that the
probability that the point $\signal$ \emph{is} covered (``hit'') by \emph{at least one} template is
\begin{equation}\label{eq:P1Nr_of_mu}
  P\left(\hit(\misnom)\, |\,\Sn,\,\Nrand\right) = 1 - \left(1 - \frac{\Vn \,\misnom^{n/2}}{\Vspace}\right)^{\Nrand}\,.
\end{equation}
This expression shows that the probability of an (unknown) signal location $\signal\in\Sn$ being covered by this random template
bank is always $< 1$ and we would require $\Nrand \rightarrow \infty$ templates to achieve certain complete coverage. However, we
can relax the requirement on
certainty of coverage and instead ask how many randomly placed templates do we need in order to obtain a probability
$\conf$ that an (unknown) signal location $\signal$ would be covered.
This is simply given by the solution to $P\left(\hit(\misnom)\, |\, \Sn, \,\Nrand\right) = \conf$, which yields
\begin{equation}\label{eq:trueNr}
  \Nrand(\conf,\,\misnom,\,\Sn) = \frac{\ln(1-\conf)}{\ln\left(1-\misnom^{n/2}\,\Vn/\Vspace\right)}\,.
\end{equation}
In practice we will mostly be interested in ``large'' parameter spaces, in the sense that the parameter-space volume $\Vspace$ is
very large compared to the volume $\Vtemplate$ of one template, and we can therefore Taylor-expand Eq.~(\ref{eq:trueNr}) in the small
quantity $\misnom^{n/2}\, \Vn/\Vspace \ll 1$, which yields
\begin{equation}\label{eq:Nr}
  \Nrand(\conf,\misnom,\Sn) \approx \frac{1}{\Vn}\ln\left(\frac{1}{1-\conf}\right) \, \misnom^{-n/2}\, \Vspace \,,
\end{equation}
where the neglected higher-order terms in the expansion correspond to corrections to $\Nrand$ of order
$\O(\ln(1-\conf)/2)$. 
We see that for a given parameter-space volume $\Vspace$, the two parameters $\misnom$ and $\conf$ completely determine the number
$\Nrand$ of random templates we need to place randomly on the parameter-space $\Sn$.
We therefore introduce the notation ${}^{\conf}\randn(\misnom)$ to denote an
$n$-dimensional random template bank with nominal mismatch $\misnom$ and covering confidence $\conf$.  An illustrative example of a $2$-dimensional
random template bank $^{0.9}\rand_2(0.1)$ is shown in Fig.~\ref{fig:example_2d_covering} for a nominal mismatch of $\misnom = 0.1$,
covering confidence $\conf=0.9$, and a parameter-space volume $\Vspace$ requiring $\Nrand=100$ templates.
By comparing Eq.~(\ref{eq:Nr}) to the general expression in Eq.~(\ref{eq:N_lattice_templates}), we can directly read
off the normalized thickness $\thickness_{\rand}(\conf)$ of a random template bank, namely
\begin{equation}\label{eq:random_thickness}
  \thickness_{\rand}(\conf) = \frac{1}{\Vn}\ln\left(\frac{1}{1-\conf}\right)\,,
\end{equation}
and from Eq.~(\ref{eq:2}) we obtain the corresponding thickness $\Thickness_{\rand}(\conf)$ as
\begin{equation}\label{eq:const_thickness}
  \Thickness_{\rand}(\conf) = \ln \left( \frac{1}{1 - \conf} \right)\,.
\end{equation}
This expression reveals a very special property of random template banks compared to any lattice covering
(e.g.\ see Eq.~(\ref{eq:cubic_thickness}) and Eq.~(\ref{eq:An_thickness})), namely the thickness $\Thickness_{\rand}$ only depends
on the covering confidence $\conf$, and is \emph{independent} of the dimension $n$.
%

\begin{figure}[tbp]
  \begin{center}
    \includegraphics[width=\columnwidth]{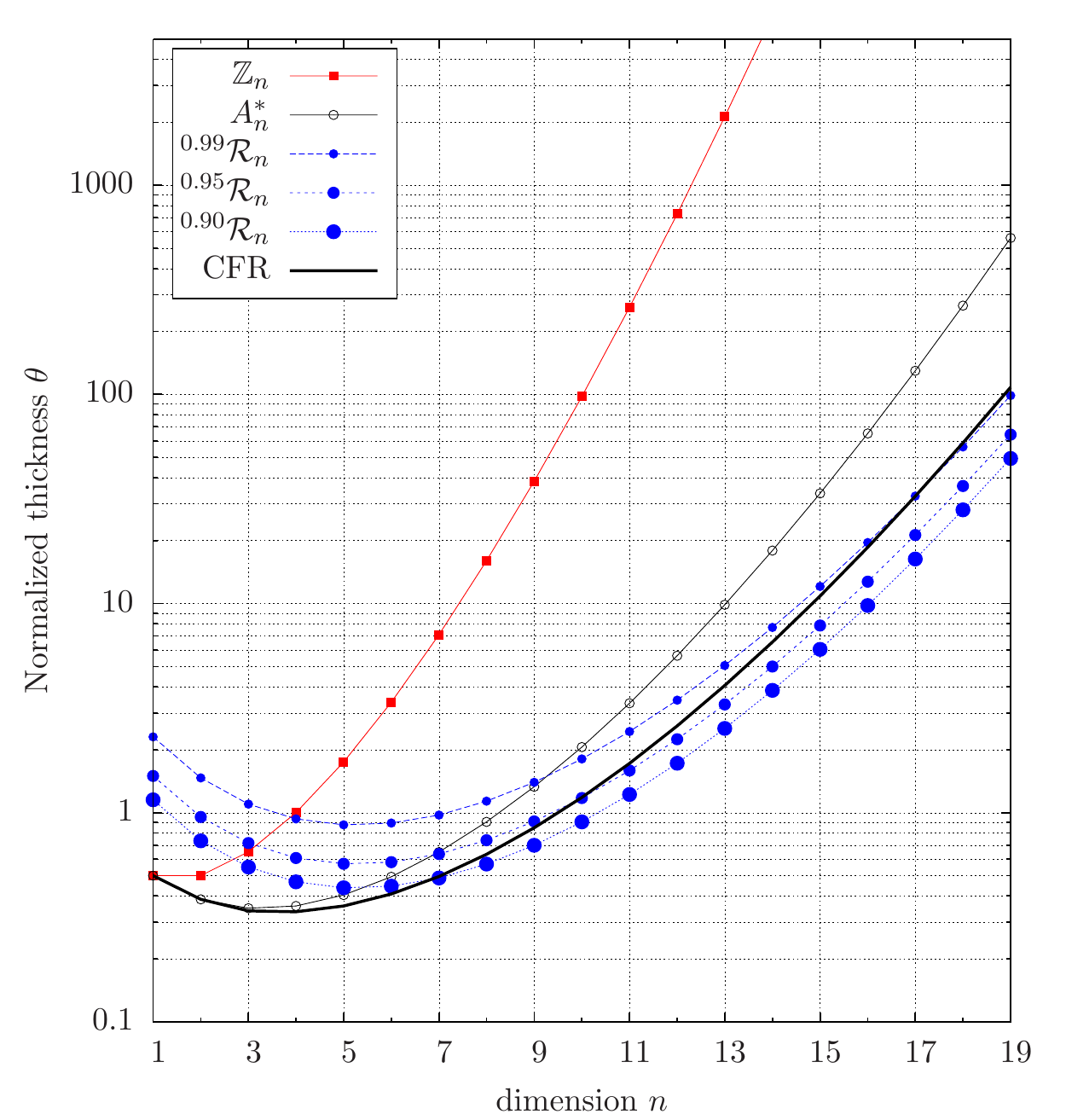}
    \caption{Normalized thickness $\thickness$ as a function of dimension $n$ for hyper-cubic ($\Zn$) and
      $\Ans$ lattice covering, and for random template banks $^{\conf}\randn$ with different choices of covering confidence
      $\conf = 0.99,\,0.95,\,0.90$. Also plotted is the Coxeter-Few-Rogers bound (CFR), which is a theoretical lower bound on the
      thickness of any strict ($\conf=1$) covering.
    }
    \label{fig:thickness}
  \end{center}
\end{figure}

Figure~\ref{fig:thickness} shows a comparison between the normalized thickness $\thickness$ of the hyper-cubic ($\Zn$) lattice
covering, the $\Ans$ lattice covering and random template banks $^{\conf}\randn$ with different covering confidences $\conf$.
We see that random template banks beat the efficiency of the $\Ans$ lattice (which is the best, or close to the best
lattice covering currently known for dimensions up to $n<24$ \cite{2007CQGra..24..481P}) at sufficiently high dimension $n$,
namely in $n=10$ for $^{0.99}\randn$, $n=7$ for $^{0.95}\randn$, and $n = 6$ for $^{0.90}\randn$. Even more surprisingly, random
template banks beat the theoretical lower bound (``Coxeter-Few-Rogers bound'' \cite{CONWAYSLOANE}) on the thickness of
\emph{any} covering! This is possible because random template banks do not provide a \emph{covering} in the strict sense,
as they leave some fraction of parameter space uncovered.

\subsection{Distribution of mismatches}
\label{sec:distr-mism}

Taking the derivative of Eq.~(\ref{eq:P1Nr_of_mu}) with respect to the nominal mismatch $\misnom$, we obtain the probability density
function (pdf) for signal mismatches $\mis$ in a random template bank of given $\Nrand$ and $\Vspace$, namely
\begin{equation}\label{eq:pdf_of_mu}
  \pdf(\mis|\Nrand,\,\Sn) = \frac{n \,\Nrand \, \Vn \, \mis^{n/2-1}}{2\Vspace}\left(1-\frac{\Vn\,\mis^{n/2}}{\Vspace}\right)^{\Nrand-1}\,,
\end{equation}
which describes the probability of finding a template within the mismatch interval
$[\mis, \mis+d\mis]$ of some location $\signal$. 
This expression is somewhat inconvenient, however, as it depends on ``extensive'' parameter-space properties,
namely the total number of templates $\Nrand$ and the parameter-space volume $\Vspace$.

In order to rewrite this purely in terms of the ``intensive'' parameters $\misnom$ and $\conf$, let us first note that for given nominal
mismatch $\misnom$ and covering confidence $\conf$, we can rewrite the required number of templates given by Eq.~(\ref{eq:Nr}) as
\begin{equation}
  \label{eq:3}
  \Nrand = \Thickness_{\rand}(\conf)\,\misnom^{-n/2}\,\No\,,\quad\mbox{with}\quad
  \No \equiv \frac{\Vspace}{\Vn}\,,
\end{equation}
where we defined $\No$ as the parameter-space volume $\Vspace$ measured in units of the unit-sphere volume $\Vn$.
Using this expression, we can write the probability of a point being covered within mismatch $\mis$ in a
random template bank ${}^{\conf}\rand(\misnom)$, using Eq.~(\ref{eq:P1Nr_of_mu}),  as
\begin{equation}
  \label{eq:6}
  P\left(\hit(\mis)\, |\, {}^{\conf}\randn(\misnom)\right) = 1 - \left(1 - \frac{\mis^{n/2}}{\No}\right)^{\No \Thickness_{\rand}\,\misnom^{-n/2}}
\end{equation}
and using the assumption of a large parameter-space volume $\Vspace$, namely $\No\gg1$, we can express
this as
\begin{equation}
  \label{eq:4}
  P\left(\hit(\mis)\,|\,{}^{\conf}\randn(\misnom) \right) \approx 1 - e^{ - \Thickness_{\rand} \, \misrelnom^{n/2}}\,,
\end{equation}
where we defined $\misrelnom$ as the mismatch $\mis$ measured in units of the nominal mismatch $\misnom$, i.e.\
\begin{equation}
  \label{eq:5}
  \misrelnom \equiv \frac{\mis}{\misnom}\,.
\end{equation}
Note that we obviously find $P\left(\hit(\misnom)\,|\,{}^{\conf}\randn(\misnom)\right) = \conf$.
Contrary to Eq.~(\ref{eq:6}), this expression for $P(\hit)$ only depends on the parameters
$\misrelnom$ and $\conf$. The interpretation of Eq.~(\ref{eq:4}) is: given a random template bank ${}^{\conf}\randn(\misnom)$
constructed for nominal mismatch $\misnom$ and covering confidence $\conf$, what is the probability of a signal location being covered
with a mismatch of at most $\mis$. By differentiating this with respect to the relative mismatch $\misrelnom$ we obtain the probability
density function of a template falling within the mismatch interval $[\misrelnom,\,\misrelnom + d\misrelnom]$ of a signal, namely
\begin{align}
  \label{eq:7}
  \pdf(\misrelnom | {}^{\conf}\randn) &\equiv \frac{d}{d\misrelnom} P\left(\hit(\mis)\,|\,{}^{\conf}\randn(\misnom)\right) \notag\\
  &= \frac{n}{2}\, \Thickness_{\rand}\,  \misrelnom^{{n}/{2} - 1} \,
  e^{- \Thickness_{\rand}\, \misrelnom^{n/2}}\,.
\end{align}
\begin{figure}
  \begin{center}
    \includegraphics[width=\columnwidth]{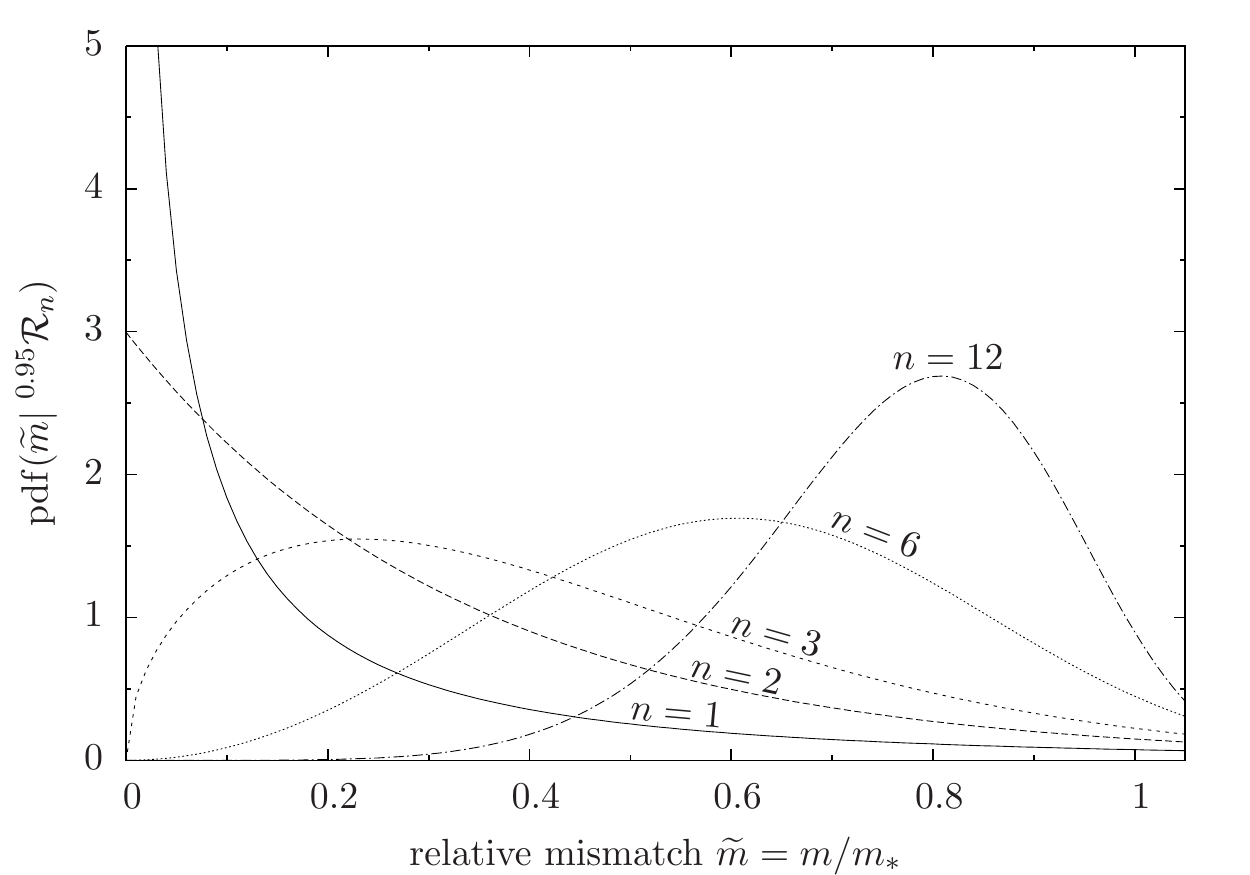}
    \caption{Probability distribution Eq.~(\ref{eq:7}) of relative mismatches $\misrelnom=\mis/\misnom$ in
      random template banks ${}^{0.95}\randn$ of dimension $n= 1, 2, 3, 6, 12$.}
    \label{fig:mismatchPDF_Rand95}
  \end{center}
\end{figure}
%
Figure~\ref{fig:mismatchPDF_Rand95} shows a plot of the mismatch pdf for random template banks $^{0.95}\randn$ in different
dimensions $n$. We can see in this plot that at higher dimensions the bulk of the mismatch probability shifts to the higher end of
relative mismatches $\misrelnom$. This shows that random template banks become more and more efficient as the number of dimensions
grows. The reason for this is that a bank that yields mismatch distributions peaked at lower values of the relative mismatch must
be more densely populated than a bank with mismatch distribution peaked at higher values of the relative mismatch. In order to see
this it is useful to stress that we construct template banks with the only constraint that the maximum mismatch be lower than a given
fixed value, $\misnom$ (with probability $\conf$): if that constraint is satisfied the bank is acceptable. We can construct
different banks that satisfy this
constraint, with different template densities. In a bank that is very densely populated the average distance to the nearest
template is smaller than in a bank with lower template density. Hence in a more densely populated bank the distribution of mismatches
peaks at a lower value of the relative mismatch compared to a lower-density bank. Among the two banks, the lower-density one
has fewer templates, hence it is more efficient. 

\subsection{Spatial parameter-space coverage}
\label{sec:spat-param-space}

In our discussion of random template banks so far we have focused on the probability $\conf$ that an (unknown) signal location
$\signal$ is covered by a template. If we were to construct a number of random template banks, this ``confidence'' therefore
describes the fraction of template banks in which $\signal$ would be covered. A somewhat related, yet different, question is:
given a realization of a random template bank $^{\conf}\randn(\misnom)$, what is the fraction $\cov$ of parameter space that is \emph{actually} covered?
This \emph{spatial coverage} $\cov$, as a property of an individual random-template-bank realization, is a random variable, and in
the following we will analyze some of its relevant statistical properties.

For a given random-template-bank realization, we define the function $f(\lambda)$ on the parameter-space $\Sn$ as
\begin{equation}\label{eq:f}
  f(\lambda) = \left\{\begin{array}{l}
      1\quad \mbox{if $\lambda$ is covered} \\
      0\quad \mbox{otherwise,}
    \end{array}\right.
\end{equation}
describing whether a point $\lambda$ is covered (``hit'') or not (``miss'') within mismatch $\misnom$.  Using this we can
express the spatial coverage fraction $\cov$ as
\begin{equation}\label{eq:true_coverage}
  \cov = \frac{1}{\Vspace}\int_{\Sn} f(\lambda) \, dV\,,
\end{equation}
where $d V = \sqrt{g}\, d^{n}\lambda$ is the volume element associated with $d^n\lambda$.
Given any point $\lambda \in \Sn$ the expectation value of $f(\lambda)$ over an ensemble of template banks is given by
\begin{align}
  E\left[f(\lambda)\right] &= \sum_{j=0}^{1} j \, P\left(f(\lambda) = j\right)\notag\\
  &= P(\hit(\misnom)\, | {}^{\conf}\randn(\misnom)) \notag\\
  &= \conf\,,
\end{align}
where we used the fact that the probability of any point $\lambda$ being covered in $^{\conf}\randn(\misnom)$ is by construction
given by the covering confidence $\conf$.  Using this we can express the expectation value of the spatial coverage $\cov$ as
\begin{equation}\label{eq:exp-cov}
  E\left[\cov\right] = \frac{1}{\Vspace} \int_{\Sn} E\left[f(\lambda)\right] \, dV = \conf\,,
\end{equation}
showing that the expected spatial coverage $\cov$ is equal to the covering confidence $\conf$. Note that this result holds true despite the
obvious existence of correlations in $f(\lambda)$ between neighboring points.

\begin{figure}
  \begin{center}
    \includegraphics[width=\columnwidth]{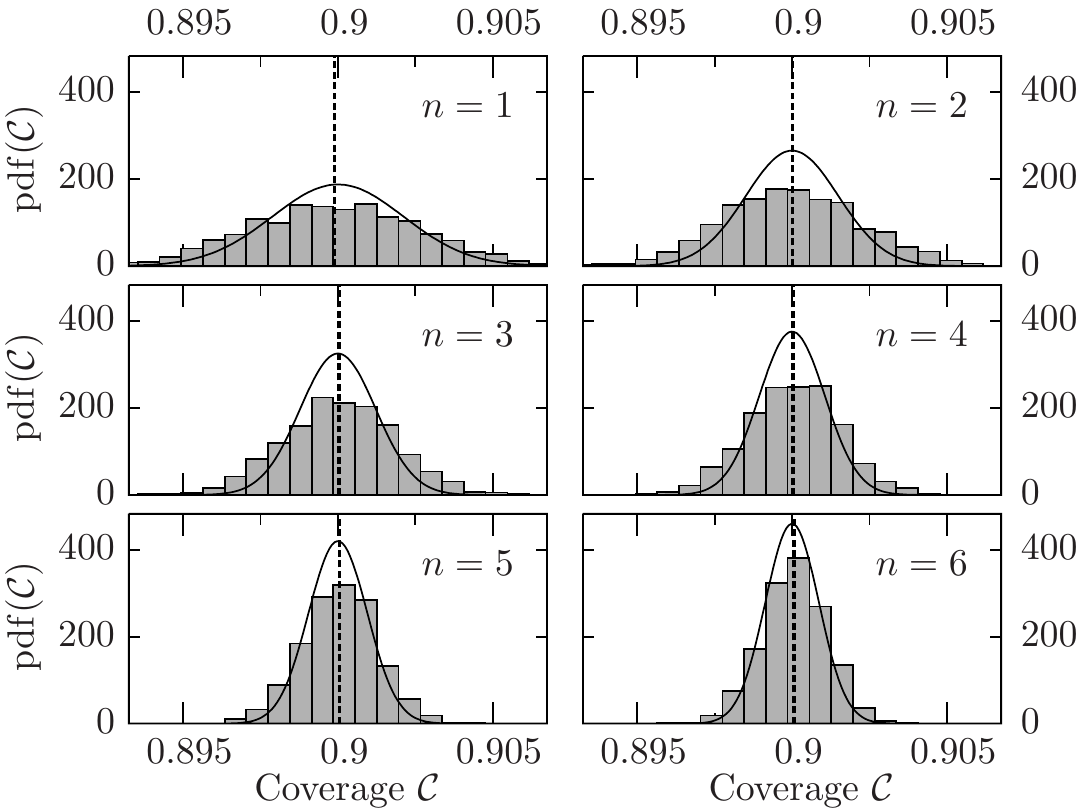}
    \caption{Results of Monte-Carlo simulations of spatial coverage $\cov$ for realizations of ${}^{0.9}\randn$ random template
      banks in $n = 1,\ldots,6$ dimensions, using $\Nrand = 10^4$ templates.
      The histograms show the distribution of measured coverages $\cov$, the solid black curves show a Gaussian distribution
      of mean $\conf$ and variance given by Eq.~(\ref{eq:var_estimate}), and the dashed black lines indicate the mean values of
      the measured distributions.
    }
    \label{fig:eta_distribution}
  \end{center}
\end{figure}

However, the question of the \emph{variance} in spatial coverage $\cov$ over an ensemble of random template banks is substantially
complicated by these spatial correlations, as we need to evaluate the following integral
\begin{equation}
  \var\left[\cov\right] = \frac{1}{\Vspace^2}\left\{ \int_{\Sn} dV \int_{\Sn} dV' \, E\left[f(\lambda)\,f(\lambda')\right]\right\}  - \conf^{2}\,.
\end{equation}
Since we have no good handle on the spatial correlations in $f(\lambda)$ for a given realization of $^{\conf}\randn$, we
try to find some reasonable approximations.  First we approximate the above integral as a discrete sum over finite volume
elements $\Delta V$. In addition we assume that $f(\lambda)$ in each volume element $\Delta V$ is completely uncorrelated with
$f(\lambda')$ in any other volume element, and that $f(\lambda)$ is perfectly correlated within each volume element.
Applying these approximations yields the following estimate for the variance:
\begin{equation}\label{eq:var-deltaV}
  \var\left[\cov\right] \approx \frac{\Delta V}{\Vspace}\,\conf \, (1-\conf)\,.
\end{equation}
We have found that a reasonably good semi-empirical ``guess'' for the number $N_\ind$ of independent ``uncorrelated''
volume elements in $\Sn$ seems to be
\begin{equation}
  \label{eq:8}
  N_\ind \sim 2 \, n \, \Nrand\,,\quad\mbox{and}\quad\Delta V = \frac{\Vspace}{N_\ind}\,.
\end{equation}
Note that we do currently not have a good theoretical understanding of this number, however, it seems to yield reasonably good
quantitative agreement with the Monte-Carlo results on the coverage, as well as the worst-case mismatch discussed in the next
section. We therefore present the analytic estimate as a useful indicator of the general behavior and trends at higher dimensions.
Substituting this into Eq.~(\ref{eq:var-deltaV}), we obtain the following rough estimate for the coverage variance
\begin{equation}\label{eq:var_estimate}
  \var[\cov] \approx \frac{\conf \, (1-\conf)}{2 \, n \, \Nrand}\,.
\end{equation}
A key feature to notice is the inverse proportionality of the variance to the number of templates $\Nrand$ and the
parameter-space dimension $n$. In Fig.~\ref{fig:eta_distribution} we show example distributions of the spatial coverage $\cov$,
obtained through numerical simulations, and compare them to Gaussian distributions of mean $\conf$ and variance given by
Eq.~(\ref{eq:var_estimate}). 
Each simulation used $2000$ random-template-bank realizations (parameter space volumes were computed as a function of $\Nrand$
and $\conf$ and edge effects removed using periodic boundary conditions).
A Monte-Carlo integration using $5\times 10^5$ points was then performed on each realization to compute the spatial coverage
$\cov$. These simulations become very computationally intensive as the dimensionality of the space
increases and hence we limited our test scenarios to $n \le 6$ and $\Nrand = 10^4$.  The Gaussian
distributions using mean $E[\cov]$ and variance $\var[\cov]$ seem to agree quite well with the results from the
Monte-Carlo simulations for the range of dimensions considered.

\subsection{Expected worst-case mismatch}
\label{sec:expected-worst-case}          

A related question to the spatial coverage $\cov$ is the worst-case mismatch $\misworst$ found in an individual realization of
${}^{\conf}\randn(\misnom)$, i.e.\ the ``deepest hole'' in the template bank.
As seen in the previous section, a fraction $1-\cov$ of the parameter space $\Sn$ will have mismatch $\mis>\misnom$. Within this
uncovered fraction there will be a worst-case location with the largest value $\misworst$ of mismatch.
Similar to $\cov$, the worst-case mismatch $\misworst$ is a random variable dependent on a given realization of
$^{\conf}\randn(\misnom)$, and in this section we analyze the statistical properties of $\misworst$ over an ensemble of random-template-bank 
realizations.

The geometrical definition of the minimal mismatch $\mis(\lambda)$ in any parameter-space point $\lambda$ for a given random
template bank can be expressed as
\begin{equation}\label{eq:mindist_function}
  \mis(\lambda) = \min^{\Nrand}_{k=1}\left\{g_{ij}\,\Delta \lambda^i_{(k)}\,\Delta \lambda^j_{(k)}\right\}\,,
\end{equation}
where $k$ indexes the $\Nrand$ random templates $\lambda_{(k)}$ and $\Delta\lambda_{(k)} = \lambda - \lambda_{(k)}$.  The
worst-case mismatch $\misworst$ on the space $\Sn$ is therefore
\begin{equation}\label{eq:worst-mismatch}
  \misworst \equiv \max_{\Sn}\left\{\mis(\lambda)\right\}\,.
\end{equation}
Note that this is in fact the definition of the ``covering mismatch'' (or covering radius, squared) of a set of templates
\cite{CONWAYSLOANE}, and for lattices and other complete ($\conf=1$) coverings, this will be equal to the nominal mismatch,
i.e.\ $\misworst = \misnom$.
In the case of random template banks, however, we control the nominal mismatch $\misnom$, but the worst-case mismatch could take
on arbitrarily large values $\misworst > \misnom$.
One very interesting approach to answering this question would be to use the fact that all ``local'' worst-case mismatches occur at
Voronoi cell vertices~\cite{Okabe00} (being the so-called ``holes'' of the covering). For a given number of random templates
(or ``Voronoi seeds'') the expectation value of the number of vertices is known~\cite{brakke:ndvoronoi}.  This would be a good
start to defining a finite set of points for which to draw random mismatch values. The problem with this approach, however, is that
the vertex density grows exponentially with dimension, and vertex mismatch values therefore will become highly correlated due to
their relative ``closeness''. In addition, while we know the mismatch distribution (see Eq.~(\ref{eq:7})) at randomly selected points
$\lambda$, the mismatch distribution on Voronoi \emph{vertices} (being very special points) is different, and a complicated function of
$\pdf(\misrelnom|{}^{\conf}\randn)$.

We therefore employ a more direct and crude approach, namely using the semi-empirical ``guess'' Eq.~(\ref{eq:8})
for the number $\Nind$ of statistically ``independent'' locations in $\Sn$, together with the known mismatch
distribution of Eq.~(\ref{eq:7}).
A linear rescaling of the whole template space will affect all mismatches equally, and therefore it will be more useful in the
following to consider the relative worst-case mismatch
$\misrelnom_\worst \equiv \misworst/\misnom$, i.e.\ $\misworst$ measured in units of the nominal mismatch $\misnom$.

\begin{figure}
  \begin{center}
    \includegraphics[width=\columnwidth]{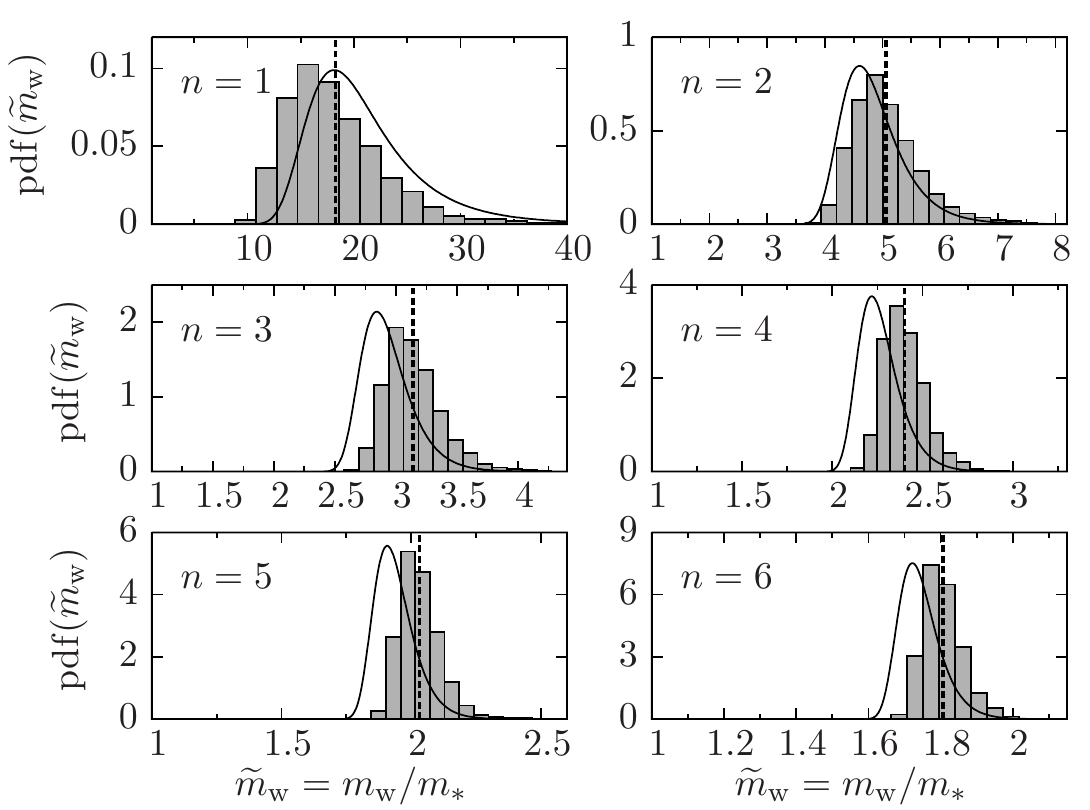}
    \caption{Results of Monte-Carlo simulations of (relative) worst-case mismatch $\misrelnom_\worst$ for
      realizations of ${}^{0.9}\randn$ random template banks in $n = 1,\ldots, 6$ dimensions,
      using $\Nrand = 10^4$ templates. The histograms show the distribution of measured (relative) worst-case mismatches
      $\misrelnom_\worst$, the solid black curves show the estimated distribution Eq.~(\ref{eq:12}) and the dashed black lines
      indicate the mean values of the measured distributions.
    }
    \label{fig:worstcase1}
  \end{center}
\end{figure}

\begin{figure}
  \begin{center}
    \includegraphics[width=\columnwidth]{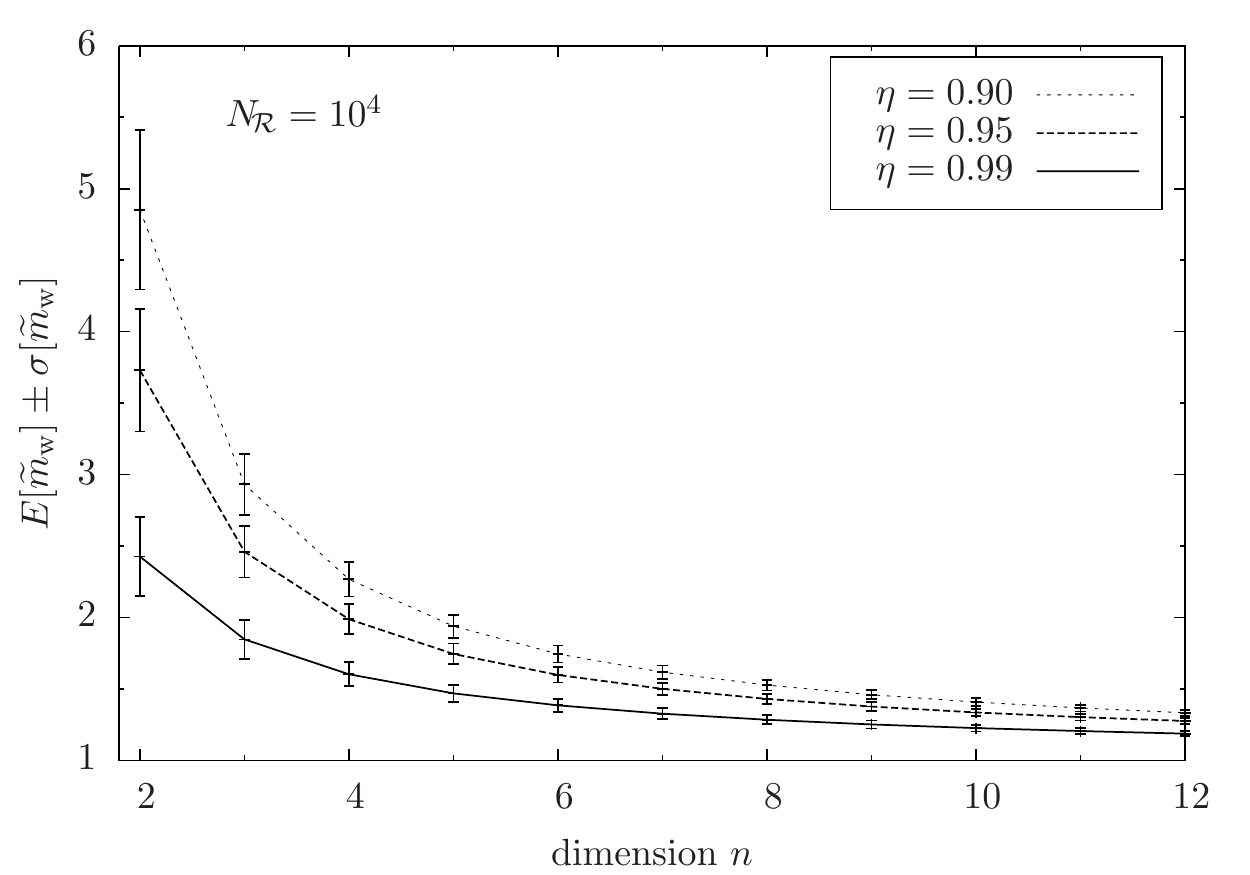}\\
    \includegraphics[width=\columnwidth]{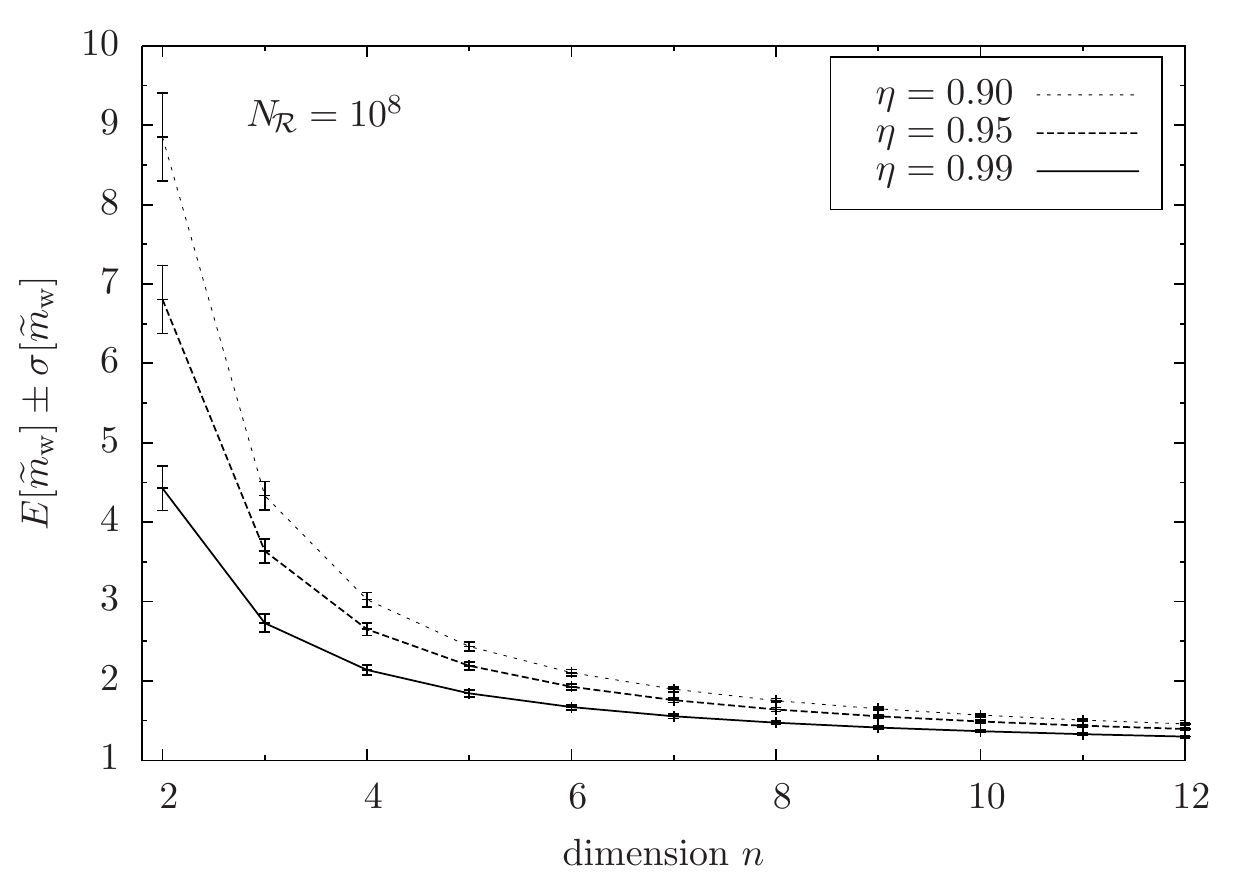}
    \caption{Statistical properties of the worst-case mismatch $\misrelnom_\worst$ in random template banks as function of
      dimension $n$, covering confidence $\conf$ and number of templates $\Nrand$.
      This plot shows the expectation value $E[\misrelnom_\worst]$ and standard
      deviation $\sigma[\misrelnom_\worst]$ derived from the theoretical estimate Eq.~(\ref{eq:12}).
      The upper panel shows the case for $\Nrand=10^4$ templates, while the lower panel is for $\Nrand =10^8$.
    }
    \label{fig:worstcase-vs-n}
  \end{center}
\end{figure}

The probability that the \emph{largest} (relative) mismatch $\misrelnom$ of $\Nind$ independent trials falls within the interval
$[\misrelnom_\worst,\,\misrelnom_\worst+d\misrelnom]$ can be expressed as
\begin{equation}
  \label{eq:9}
  \pdf(\misrelnom_\worst|\conf,\,\Nind)\,d\misrelnom = \binom{\Nind}{1}\,p_1\,p_0^{\Nind-1}\,,
\end{equation}
where $p_1$ is the probability of the mismatch in a single trial falling within
$[\misrelnom_\worst,\,\misrelnom_\worst+d\misrelnom]$, i.e.\
\begin{equation}
  \label{eq:10}
  p_1 \equiv \pdf(\misrelnom_\worst|\,{}^\conf\randn)\,d\misrelnom\,,
\end{equation}
and $p_0$ is the probability of the mismatch in a single trial falling within $[0,\,\misrelnom_\worst]$, i.e.\
\begin{equation}
  \label{eq:11}
  p_0 \equiv P\left(\hit(\mis_\worst)\, | \,{}^{\conf}\randn(\misnom)\right)\,.
\end{equation}
Using Eq.~(\ref{eq:7}) this can be rewritten as
\begin{equation}
  \label{eq:12}
  \begin{split}
    &\pdf(\misrelnom_\worst|\,{}^{\conf}\randn,\,\Nind) = \left. \frac{d}{d \misrelnom}\right|_{\misrelnom_\worst} P(\hit|\misrelnom,\,\conf)^{\Nind}\\
    &= \frac{n}{2}\Thickness_\rand \Nind \misrelnom_\worst^{n/2-1}e^{-\Thickness_\rand\misrelnom_\worst^{n/2}}
    \left( 1 - e^{-\Thickness_\rand \misrelnom_\worst^{n/2}}\right)^{\Nind-1}\,.
  \end{split}
\end{equation}
Using this expression we can easily compute quantiles of the worst-case mismatch distribution, namely from
\begin{align}
  \label{eq:13}
  P(\misrelnom_\worst \le \misrelnom |\conf,\,\Nind) &= \int_0^{\misrelnom} \,
  \pdf(\misrelnom_\worst|\conf,\,\Nind)\,d\misrelnom_\worst \notag\\
  &= \left( 1 - e^{\Thickness_{\rand} \misrelnom^{n/2}} \right)^{\Nind}\,,
\end{align}
so for example we obtain the median worst-case mismatch as
\begin{equation}
  \label{eq:14}
  \misrelnom_\worst^{50\%} = \left[ - \frac{1}{\Thickness_{\rand}} \ln \left( 1 - 2^{-1/\Nind}\right)\right]^{2/n} \,.
\end{equation}
Using the semi-empirical guess, given in Eq.~(\ref{eq:8}), for the number $\Nind$ of ``independent'' parameter-space points in $\Sn$, we
obtain quantitative estimates for the distribution of the worst-case mismatch.
Figure~\ref{fig:worstcase1} shows a few example worst-case mismatch distributions, comparing the analytical estimate of Eq.~\eqref{eq:12}
to the results of Monte-Carlo simulations. These simulations were run with 2000 random-template-bank realizations of a 
${}^{0.9}\randn$ bank containing $\Nrand = 10^4$ templates each.
The predicted distributions differ slightly from the simulations, but this should be expected given our crude estimation of
$\Nind$. With the exception of $n=1$, the analytical expression tends to slightly underestimate the worst-case mismatch values.
Nevertheless, our rough estimate seems to capture the overall trend to smaller values of $E[\misrelnom_\worst]$ and
$\var[\misrelnom_\worst]$ with increasing dimension $n$.
This behavior is also illustrated in Fig.~\ref{fig:worstcase-vs-n}, where we have plotted the analytic expectation value
and variance (computed from Eq.~(\ref{eq:12})) of $\misrelnom_\worst$ as a function of dimension $n$ for various covering confidences $\conf$
and numbers of templates $\Nrand$.
We see that in higher dimensions the expectation value of the relative worst-case mismatch asymptotically approaches unity.  We also see that as the
number of templates $\Nrand$ increases, corresponding to a larger number of mismatch ``trials'', we obtain an increase in the
expected value of $\misrelnom_{\worst}$.
In addition, we see that the standard deviation $\sigma[\misrelnom_{\worst}] \equiv \sqrt{\var[\misrelnom_\worst]}$ decreases both with increasing
 number of template $\Nrand$ as well as with higher dimension $n$, so that worst-case mismatches become both smaller and more
tightly constrained in these limits.  For example, as shown in Fig.~\ref{fig:worstcase1}, for $n=6$ our Monte-Carlo
simulations give the mean value of $\misrelnom_{\worst}$ as $1.81$ with a standard deviation of $0.06$ for a random template
bank containing $\Nrand=10^4$ templates with a covering confidence $\conf=0.9$.  In using such a template bank one would therefore
expect with $> 99\%$ ($3\sigma$) confidence that the largest mismatch is smaller than $1.99\,\misnom$.
We should also note that despite our rather crude estimate for the number of independent parameter space locations $\Nind$
our model is able to estimate the mean value of $\misrelnom_{\worst}$ to within $\approx 10\%$ of the values obtained from our
simulations.
This is in fact one of the key results from this investigation, that although random template banks by construction only provide incomplete
coverage, $\cov < 1$, the actual worst-case mismatch $\misworst$ in the uncovered regions is not expected to be substantially
larger than the nominal mismatch $\misnom$, especially at higher dimensions.

%
\subsection{Practical issues in curved parameter-spaces}
\label{sec:nonflat-spaces}
%
As noted in the beginning of the section, the random template bank results apply without modification to curved parameter
spaces and non-constant metrics.  In order to construct such a template bank in practice, however, we need to
generate a uniform random sampling \emph{in proper volume}, which for non-constant metrics can be non-trivial.
In order to see how to take account of non-constant metric components $g_{ij}(\lambda)$ in the random template placement, we note
that Eq.~\eqref{eq:Nr} specifies a constant uniform probability density $\dens_{\rand}$ of templates, namely
\begin{equation}
  \label{eq:16}
  \dens_{\rand} = \frac{\Nrand}{\Vspace} = \frac{d\Nrand}{d V} = \frac{1}{\sqrt{g}}\,\frac{d \Nrand}{d^n\lambda}\,.
\end{equation}
In other words, the non-constant template ``pseudo-density'' $\denscoord_{\rand}(\lambda)$ in \emph{coordinate-space} satisfies
\begin{equation}
  \label{eq:17}
  \denscoord_{\rand}(\lambda) \equiv \frac{d\Nrand}{d^n\lambda} = \sqrt{g(\lambda)}\,\dens_{\rand} = \sqrt{g(\lambda)}\,\thickness_{\rand}\,\misnom^{-n/2}\,,
\end{equation}
which specifies the required random sampling density in coordinate space.

There are various sophisticated and efficient methods for sampling from non-uniform distributions, Markov-Chain-Monte-Carlo
(MCMC) methods, importance resampling, and rejection sampling are a few examples.  In each of these sampling methods it is sufficient 
to know the density $\denscoord_{\rand}(\lambda)$ (and hence the metric determinant) only up to some normalising constant factor.  
However, one must know $\Nrand$, the total number of random templates to draw, which requires accurate knowledge of the proper volume, 
defined by Eq.~\ref{eq:proper_volume}, and therefore ultimately one must also have accurate knowledge of the metric determinant.

The simplest method, applicable for slowly varying template densities
is to decompose the parameter space $\Sn$ into smaller patches $\Sn^{(j)}$, which are small enough so they can be approximated by a
constant metric, and sampling it uniformly with template density given by Eq.~(\ref{eq:17}) evaluated at the centre of each patch.

\section{Relaxed lattices}  
\label{sec:relaxed-lattices}
%
In the previous section we saw that random template banks ${}^{\conf}\randn$
will outperform \emph{any} covering at sufficiently high parameter-space dimension. One of the
key features of random template banks, however, is that they do not actually provide a strict covering: any point is only covered with
probability $\conf < 1$.
This allows random template banks to beat even the theoretical (Coxeter-Few-Rogers) lower bound on the thickness of coverings.
In higher dimensions it seems to get extremely expensive (in terms of number of templates) to cover the ``last few percent''
of a parameter space. Relaxing the requirement of complete coverage therefore allows enormous
gains in efficiency.

We can now apply this insight to lattice coverings, by relaxing the strict ``minimax'' prescription and instead requiring a
mismatch $\misnom$ only with probability $\conf < 1$. This allows us to use a larger maximal mismatch
$\mismax > \misnom$ for the $n$-dimensional covering lattice $\Latticen(\mismax)$, thereby reducing the required number
of templates. The relation between these quantities is given by
\begin{equation}\label{eq:relaxed-confidence}
  \conf = \int_0^{\misnom} \pdf \left( \mis | \,\Latticen(\mismax) \right)\, d\mis\,,
\end{equation}
where $\pdf\left( \mis | \,\Latticen(\mismax)\right)$ is the probability distribution of mismatches $\mis$ for sampled
points within a lattice $\Latticen(\mismax)$. A uniform linear rescaling of the whole template space will affect
all mismatches equally, and therefore it will be more useful to introduce the relative mismatch
$\misrelmax \equiv \mis / \mismax$, which is invariant under re-scalings.
Note that contrary to Sec.~\ref{sec:random-template-placement}, here we use a relative mismatch defined with respect to
$\mismax$ of the lattice, while the nominal mismatch $\misnom$ is a-priori unknown and will be determined from
Eq.~(\ref{eq:relaxed-confidence}). In Sec.~\ref{sec:random-template-placement} a relative mismatch
$\misrelnom \equiv \mis/\misnom$ was used, because there was no strict maximal mismatch in this case and we directly prescribed
the nominal mismatch $\misnom$.
The probability distribution of $\misrelmax$ is $\pdf(\misrelmax | \Latticen) = \mismax\, \pdf(\mis|\Latticen(\mismax))$, such that
$\int_0^1 \pdf(\misrelmax|\Latticen)\,d\misrelmax = 1$. We can therefore restate Eq.~(\ref{eq:relaxed-confidence}) in the more useful
form
\begin{equation}\label{eq:relaxed-relative-confidence}
  \conf = \int_0^{\miseffrel} \pdf(\misrelmax | \Latticen)\, d\misrelmax\,,
\end{equation}
where we defined the ``effective'' relative mismatch $\miseffrel \equiv \misnom/\mismax<1$, which is determined by the lattice mismatch distribution
$\pdf(\misrelmax|\Latticen)$ and the covering confidence $\conf$.  We define the linear \emph{relaxation factor}
$\linrelax$ for the relaxed lattice ${}^\conf\Latticen$ as
\begin{equation}\label{eq:relaxation-factor}
  \linrelax\left({}^{\conf}\Latticen\right) \equiv \sqrt{\frac{\mismax}{\misnom}} = \miseffrel^{-1/2} > 1\,.
\end{equation}
For a given nominal mismatch $\misnom$ and covering confidence $\conf$, the relaxation factor determines the maximal covering
mismatch $\mismax$ of the lattice as
\begin{equation}
  \label{eq:18}
  \mismax = \misnom \, \linrelax^2\left({}^{\conf}\Latticen\right)\,,
\end{equation}
which defines the ``relaxed lattice'' as ${}^{\conf}\Lattice(\misnom) \equiv \Lattice(\mismax)$.
The normalized thickness $\thickness_{\Latticen}(\conf)$
of a relaxed lattice ${}^{\conf}\Latticen$ is thereby reduced to
\begin{equation}\label{eq:relaxed-thickness}
  \thickness_{\Latticen}(\conf) = \frac{\thickness_{\Latticen}}{r^n\left({}^{\conf}\Latticen\right)}\,,
\end{equation}
with respect to the thickness $\thickness_{\Latticen} = \thickness_{\Latticen}(\conf=1)$ of a traditional covering lattice.
Note that contrary to random template banks, the spatial coverage fraction $\cov$ of relaxed lattices is not a random variable
and corresponds exactly to the covering confidence, i.e.\ $\cov = \conf$ and $\var[\cov] = 0$.
Furthermore, the worst-case mismatch is also exactly known, namely $\mis_\worst = \mismax$.

%
%
As seen in Eq.~(\ref{eq:relaxed-relative-confidence}) and Eq.~(\ref{eq:relaxation-factor}), the relaxation factor
$\linrelax({}^{\conf}\Latticen)$ is determined from the mismatch distribution $\pdf(\misrelmax|\Latticen)$. The distribution of
mismatches depends strongly on the type of lattice $\Latticen$ and the dimension $n$.  Unfortunately, these mismatch distributions
are generally not known analytically, and we need to resort to Monte-Carlo simulations to determine them.

\begin{figure}
  \begin{center}
    \includegraphics[width=\columnwidth]{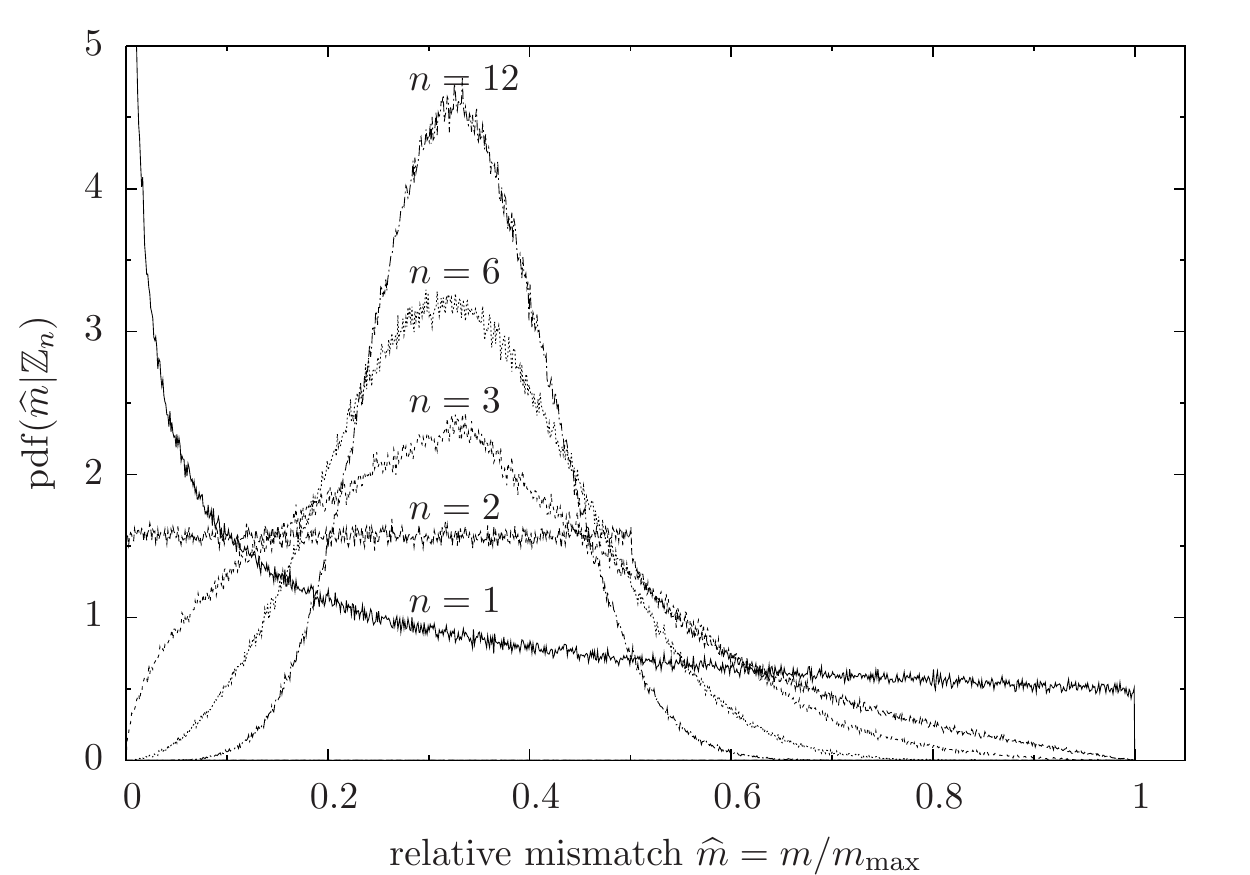}
    \caption{Results of Monte-Carlo simulations (each using $10^6$ points) of the distribution of relative mismatches $\misrelmax$
      in hyper-cubic ($\Zn$) lattices in dimensions $n= 1,\,2,\,3,\,6,\,12$.
    }
    \label{fig:mismatchPDFZn}
  \end{center}
\end{figure}

\begin{figure}
  \begin{center}
    \includegraphics[width=\columnwidth]{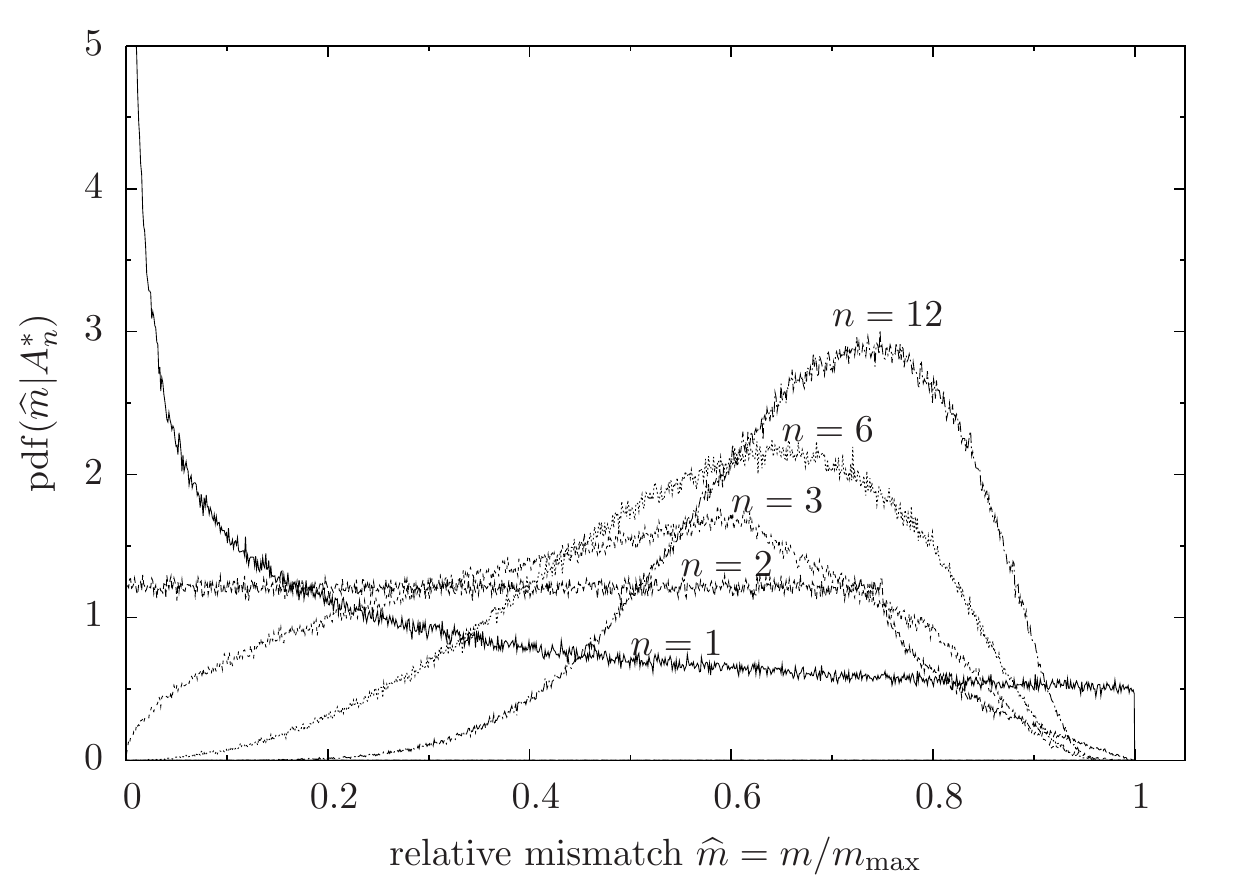}
    \caption{Results of Monte-Carlo simulations (each using $10^6$ points) of the distribution of relative mismatches $\misrelmax$
      in $\Ans$ lattices in dimensions $n= 1,\,2,\,3,\,6,\,12$.
    }
    \label{fig:mismatchPDFAns}
  \end{center}
\end{figure}

In order to sample the lattice mismatch distribution, we uniformly pick points in parameter space, then find the closest lattice
template and determine its relative mismatch $\misrelmax=\mis/\mismax$.
For finding the closest template, we use an elegant and efficient method described in \cite{conway:Voronoi1984}, which is based on
the ``fast quantizing'' algorithms available for many well-known lattices (see \cite{conway:Quantizing1982} and
\cite{CONWAYSLOANE} Chapter~20). Here we focus on two lattices only: the simple but inefficient hyper-cubic lattice
$\Zn$, and the highly efficient covering lattice $\Ans$.  Some resulting sampled mismatch distributions are
shown in Figs.~\ref{fig:mismatchPDFZn} and \ref{fig:mismatchPDFAns} for different dimensions $n$.
The Monte-Carlo simulations used $10^6$ sampling points for each lattice $\Latticen$ (except for $\Ans$ in $n=18,19$, where $10^5$
points were used), and the mismatches were binned into $1000$ mismatch bins.
The ``jitter'' seen in Figs.~\ref{fig:mismatchPDFZn} and \ref{fig:mismatchPDFAns} illustrates the intrinsic sampling fluctuation
in these simulations.

\begin{figure}
  \centering
  \includegraphics[width=\columnwidth]{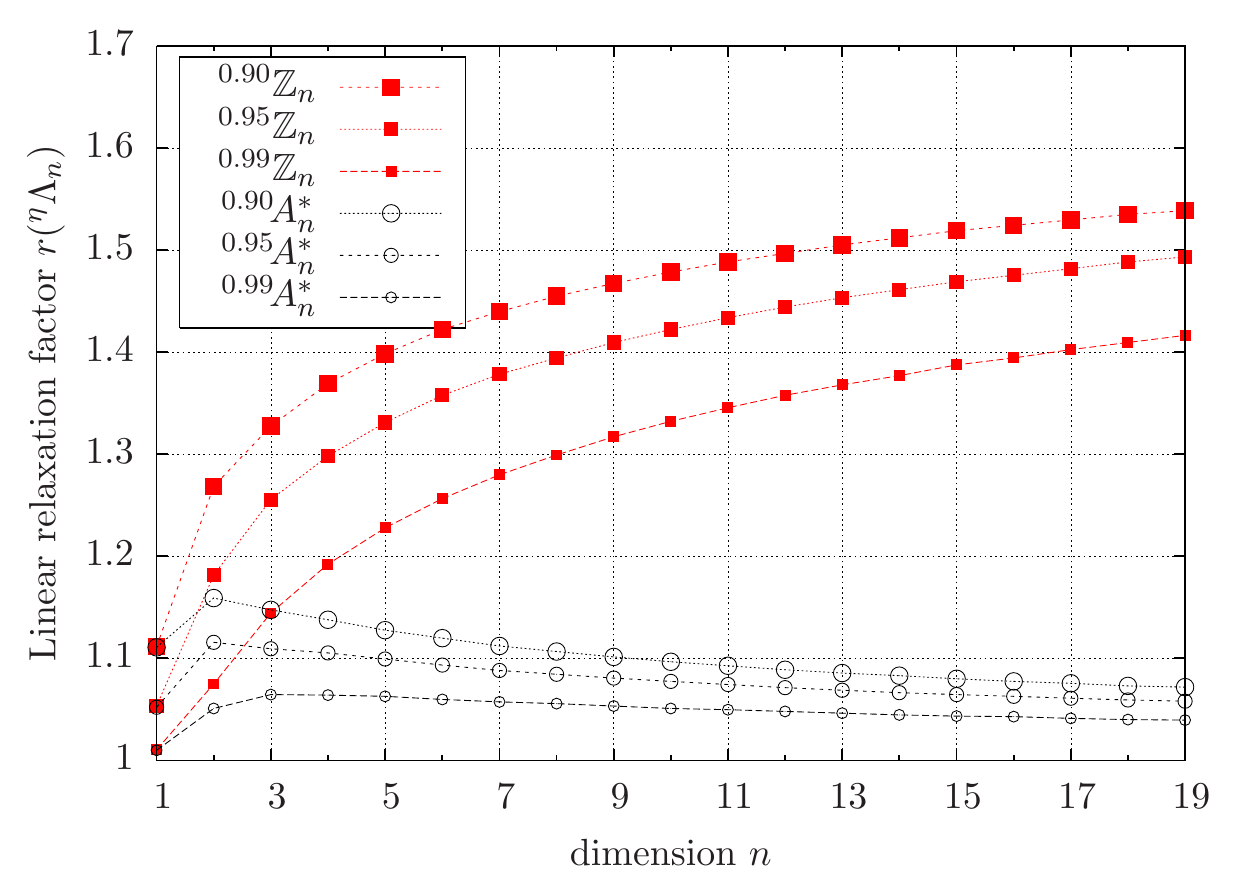}
  \caption{Linear relaxation factors $\linrelax({}^{\conf}\Lattice)$ for $\Zn$ and $\Ans$ lattices and covering confidence $\conf = 0.90,\,0.95,\,0.99$.}
  \label{fig:LinRelaxLattices}
\end{figure}

The resulting relaxation factors $\linrelax\left({}^{\conf}\Latticen\right)$ for covering confidences $\conf = 0.99,\,0.95,\,0.90$
obtained via Eq.~(\ref{eq:relaxation-factor}) are shown in Fig.~\ref{fig:LinRelaxLattices}.
The errors on the relaxation factors were determined using a Jackknife estimator (see \cite{conway:Voronoi1984}, using $100$ subsets)
and are found to be below $0.04\%$ in all cases.
We see in Fig.~\ref{fig:LinRelaxLattices} that the hypercubic ($\Zn$) lattice can be relaxed substantially more than the $\Ans$
lattice, which is also apparent from the pdfs in Fig.~\ref{fig:mismatchPDFZn} and \ref{fig:mismatchPDFAns}: the mismatch
distribution of $\Zn$ is much more ``wasteful'', as it increasingly concentrates around $\misrelmax = 1/3$. The $\Ans$ lattice on the
other hand, which is a highly efficient covering lattice, has the bulk of mismatches concentrated closer to the maximal
mismatch $\misrelmax = 1$.

\begin{figure}
  \begin{center}
    \includegraphics[width=\columnwidth]{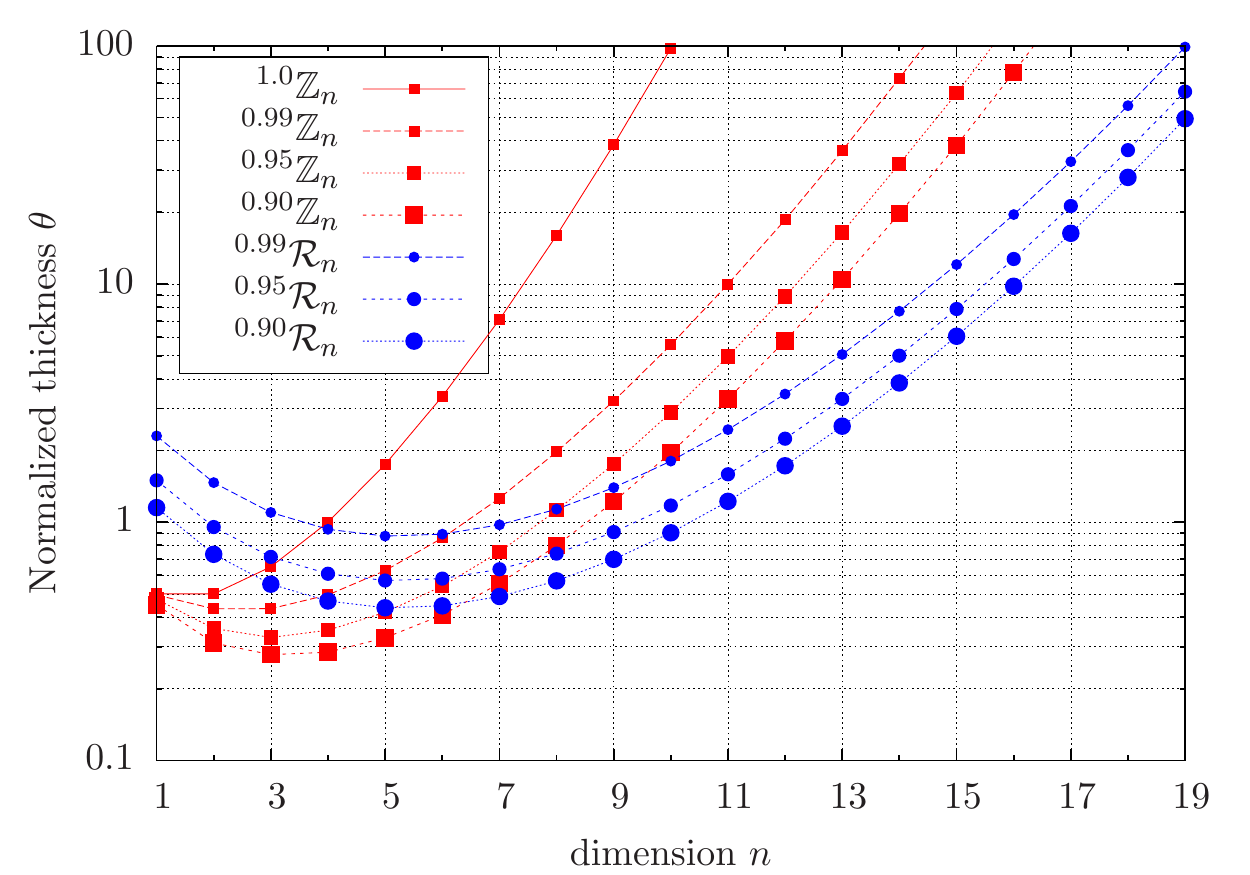}\\
    \includegraphics[width=\columnwidth]{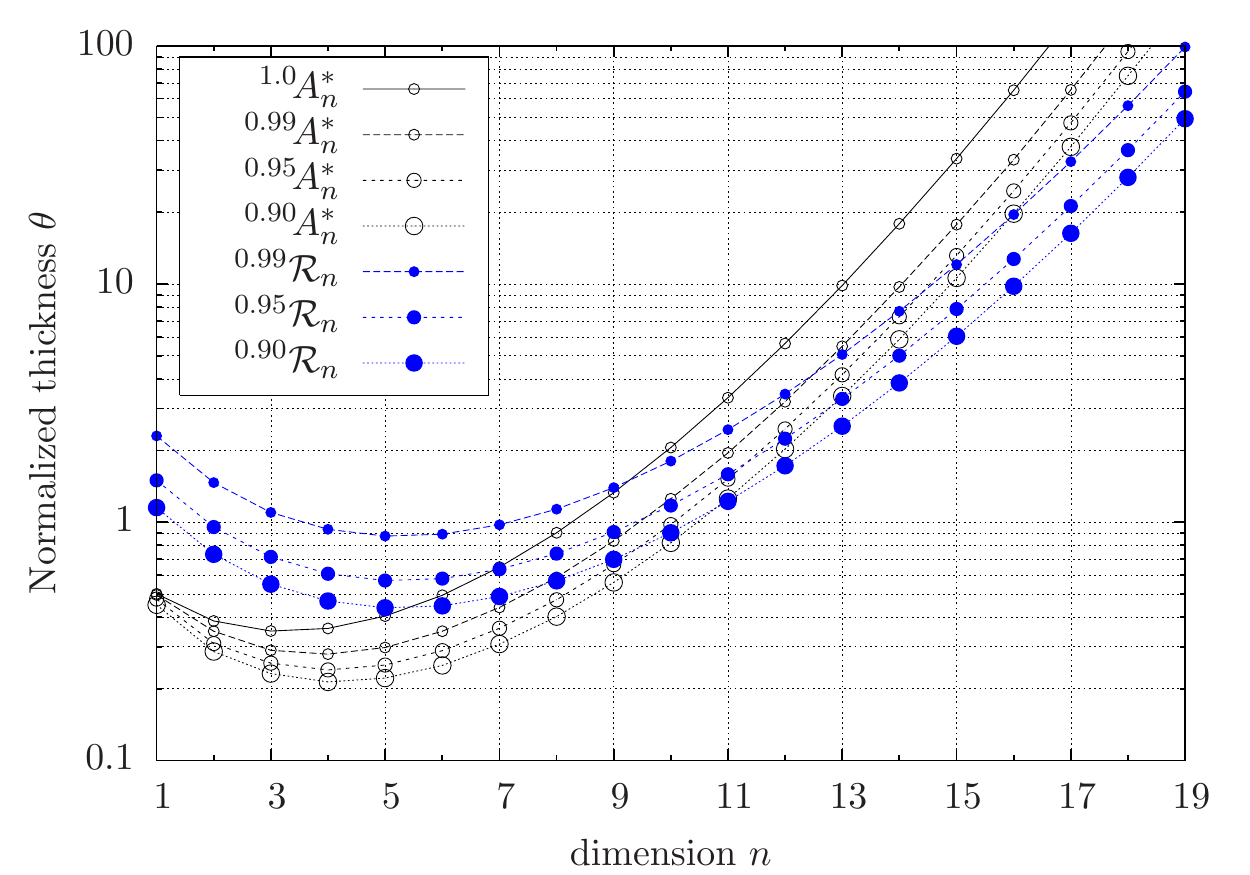}
    \caption{Normalized thickness $\thickness$  as function of dimension $n$ for strict lattice covering $\Latticen$,
      relaxed lattices ${}^{\conf}\Latticen$ and random template banks ${}^{\conf}\randn$ for values $\conf = 0.99, 0.95,
      0.90$ of covering confidence. The upper panel is for the $\Zn$ lattice, while the lower panel shows the case of the $\Ans$
      lattice.
    }\label{fig:RelaxedRandomThickness}
  \end{center}
\end{figure}

In Fig.~\ref{fig:RelaxedRandomThickness} and Table~\ref{tab:thicknesses} we show the resulting covering thickness $\thickness_{\Latticen}(\conf)$
of the relaxed ${}^{\conf}\Zn$ and ${}^{\conf}\Ans$ lattices in comparison to random template banks ${}^{\conf}\randn$.
\begin{table*}
  \begin{center}
    \input{Thickness_table.tex}
    \caption{Normalized thickness $\thickness$ in dimensions $n\leq 19$ for traditional lattice covering (${}^{1.0}\Zn$,
      ${}^{1.0}\Ans$), relaxed lattice covering (${}^{\conf}\Zn$, ${}^{\conf}\Ans$) and random template banks (${}^{\conf}\randn$)
      with covering confidences $\conf=0.99,\,0.95,\,0.90$ respectively.
      Boldface indicates the lowest thickness at given covering confidence $\conf$ and dimension $n$.
    }
    \label{tab:thicknesses}
  \end{center}
\end{table*}
We see that while relaxed lattices are substantially more efficient than traditional complete-coverage lattices, at higher
dimensions the random template banks eventually still outperform them.
At low dimensions, $n \lesssim 10$ say, the relaxed ${}^{\conf}\Ans$ lattice provides the most efficient covering we have found so
far. However, having studied only relaxed $\Zn$ and $\Ans$ lattices so far, it is conceivable that other lattices, while not
necessarily very good covering lattices, could provide an even better relaxed lattice than ${}^\conf\Ans$. More work is required
to study this possibility.

%
\section{Discussion}  
\label{sec:discussion}

Our results show that giving up deterministic certainty of coverage of a parameter space can result in large gains
in efficiency, by substantially reducing the number of required templates.
The prime example of such a \emph{relaxed covering} is the \emph{random template bank} construction, which we defined as templates placed randomly
with uniform probability distribution (per proper volume) over the parameter space.
Such a random template bank ${}^{\conf}\randn(\misnom)$ covers any signal location (excluding boundary effects) within mismatch
$\misnom$ with probability $\conf < 1$.
We have found that the template density of these random template banks can be significantly lower than that of even the most
efficient (complete) covering, and that this advantage increases in higher parameter-space dimensions.  

The exclusion of boundary
effects in the random template bank analysis is valid for situations in which the projected length scale of the templates in each of the search 
dimensions is $\ll$ the width of the parameter space in the corresponding dimension.  We should note that boundary effects have also 
been excluded in the calculation for the lattice coverings (traditional and relaxed) and in practice boundary effects will be equally
problematic for all template bank strategies.

Other studies have recently started to investigate a somewhat different random template placement strategy, referred to as
``stochastic'' template banks~\cite{babak-2008,harry-2008,harry-2008b,vdB2008:_stochastic}.
The key difference of these methods is that they involve a ``pruning stage'' in the random template placement, which is
aimed to remove templates that are too close to each other. Albeit not yet completely quantified, these methods could potentially
produce even more efficient template banks for equivalent coverage. The disadvantage, however, is that the template placement
procedure is more complicated and more computationally intensive, requiring the comparison of each template with every other.
Stochastic template banks might therefore be less suitable for higher dimensions and larger numbers of templates. In addition, it seems
unclear how much efficiency can be gained by such a pruning step, especially in higher dimensions.

Relaxed coverings do not provide complete coverage of the template parameter space.  Applying such a scheme
therefore affects the final results as an additional uncertainty. For gravitational-wave searches
such uncertainty can be well maintained at a level comparable to other uncertainties of the problem -- typically of the order of a
few percent -- hence not significantly affecting the overall degree of confidence of the result.

In computationally limited searches, relaxed template banks allow significant reductions in computational cost.
Through the re-investment of this saved computational cost, this can yield an increase in sensitivity and breadth of the search.

We have investigated some of the relevant statistical properties of random template banks, in particular the \emph{spatial}
parameter-space coverage fraction $\cov$
and the worst-case mismatch $\misworst$ expected in individual template-bank realizations.
We have performed Monte-Carlo simulations to determine the statistical distribution of these quantities, and we have found a
rough analytical estimate, which shows reasonably good agreement with the numerical results.
These results show that the variance of the spatial coverage fraction $\cov$ is inversely proportional to both
the dimensionality of the space and the number of random templates. More importantly, the worst-case mismatch $\misworst > \misnom$ is found
to be typically of the same order of magnitude as the prescribed nominal mismatch $\misnom$, and is rapidly approaching $\misnom$ with
increasing dimension $n$. At $n= 4$, for example, the largest mismatch for $\Nrand=10^8$
templates and $\eta=0.9$ has an expected value of $\sim 3\,\misnom$ with a standard deviation of $0.09\,\misnom$ (see Fig.~\ref{fig:worstcase-vs-n}).
At $n=12$, the distribution peaks at even lower values, with an expectation value of $\sim 1.5\,\misnom$ and a low standard
deviation of $0.014\,\misnom$.
Even though the coverage at the nominal mismatch $\misnom$ only holds in a statistical sense defined by the confidence $\eta$, in
practice, at dimensions greater than $n=11$, the worst-case loss is within a factor of $1.5$ of the nominal value, with very high confidence.

Inspired by these results, we have also investigated the properties of ``relaxed'' lattice coverings, which follow the
analogous prescription of a nominal covering mismatch $\misnom$, achieved with probability $\conf<1$ for any signal location.
This leads to lattices with larger maximal mismatch $\mismax > \misnom$, thereby reducing the required number of templates.
We have analyzed the properties of relaxed $\Zn$ and $\Ans$ lattices, and we have found that the relaxed ${}^{\conf}\Ans$
lattice provides the most efficient covering found so far for $n \lesssim 10$, for covering confidences $\conf \ge 0.90$.
In higher dimensions, however, random template banks outperform any other method considered so far, including
relaxed lattice coverings.

Possibly the greatest advantage of random template banks for ``real-world'' applications, however, is their practical simplicity.
Constructing lattice template banks is notoriously difficult, especially regarding the handling of curved parameter spaces and
non-constant metrics. Random template banks, on the other hand, are nearly trivial to construct, even in spaces with non-constant
metrics: one only needs to adjust the spatial probability density according to the metric determinant.

\begin{acknowledgments}
  We would like to thank Bruce Allen, Stephen Fairhurst, Graham Woan, Christian Roever, Karl Wette, Matthew Pitkin, John Veitch,
  Holger Pletsch, and the LIGO Scientific Collaboration continuous waves working group, for many useful discussions.
\end{acknowledgments}

\bibliography{randomtemplates_paper}

\end{document}

%% file: defs.tex
\newcommand{\mis}{m}
\newcommand{\misnom}{\mis_*}		
\newcommand{\mismax}{\mis_{\mathrm{\max}}}
\newcommand{\worst}{\mathrm{w}}		
\newcommand{\misworst}{\mis_\worst}	
\newcommand{\misrelmax}{\widehat{\mis}}	
\newcommand{\misrelnom}{\widetilde{\mis}}
\newcommand{\miseffrel}{\misrelmax_*}	
\newcommand{\conf}{\eta}                
\newcommand{\cov}{\mathcal{C}}          
\newcommand{\Thickness}{\Theta}		
\newcommand{\thickness}{\theta}		
\newcommand{\Ans}{A_n^*}		
\newcommand{\Zn}{\mathbb{Z}_n}		
\newcommand{\rand}{\mathcal{R}}		
\newcommand{\randn}{\rand_n}		
\newcommand{\Nrand}{N_{\!\rand}}	
\newcommand{\ind}{\mathrm{I}}		
\newcommand{\Nind}{N_\ind}      	
\newcommand{\Lattice}{\Lambda}		
\newcommand{\Latticen}{\Lattice_n}	
\newcommand{\linrelax}{r}		

\newcommand{\mcS}{\mathcal{S}}		
\newcommand{\Sn}{\mcS_n}		
\newcommand{\Vspace}{V_{\Sn}}		

\newcommand{\Vlattice}{V_{\Lattice}}	

\newcommand{\Vn}{V_n}			
\newcommand{\Vtemplate}{V_\mathrm{T}}	
\newcommand{\miss}{\,\mathrm{miss}}   
\newcommand{\hit}{\,\mathrm{hit}}     

\renewcommand{\O}{\mathcal{O}}
\newcommand{\sig}{\mathrm{s}}		
\newcommand{\signal}{\lambda_\sig} 	

\newcommand{\No}{\mathcal{N}_0}		
\newcommand{\pdf}{\mathrm{pdf}}		
\newcommand{\var}{\mathrm{Var}}		

\newcommand{\dens}{\rho}		
\newcommand{\denscoord}{\widehat{\dens}}

%% file: tag.tex
\def\commitID{commitID: 4301865fd72d5bca56a3cc57f80b6c3af5e0c117}
\def\commitDATE{Tue Sep 30 17:14:05 2008 +0200}

%% file: Thickness_table.tex

\begin{tabular}{c||ccccccccccccccccccc}
$n$ & \parbox{0.72cm}{$1$} & \parbox{0.72cm}{$2$} & \parbox{0.72cm}{$3$} & \parbox{0.72cm}{$4$} & \parbox{0.72cm}{$5$} & \parbox{0.72cm}{$6$} & \parbox{0.72cm}{$7$} & \parbox{0.72cm}{$8$} & \parbox{0.72cm}{$9$} & \parbox{0.72cm}{$10$} & \parbox{0.72cm}{$11$} & \parbox{0.72cm}{$12$} & \parbox{0.72cm}{$13$} & \parbox{0.72cm}{$14$} & \parbox{0.72cm}{$15$} & \parbox{0.72cm}{$16$} & \parbox{0.72cm}{$17$} & \parbox{0.72cm}{$18$} & \parbox{0.72cm}{$19$} \\ \hline\hline
${}^{1.0}\Zn$  & $\mathbf{0.50}$  & ${0.50}$  & ${0.65}$  & ${1.0}$  & ${1.7}$  & ${3.4}$  & ${7.1}$  & ${16}$  & ${38}$  & ${98}$  & ${261}$  & ${729}$  & ${2\mathrm{e}{3}}$  & ${6\mathrm{e}{3}}$  & ${2\mathrm{e}{4}}$  & ${7\mathrm{e}{4}}$  & ${2\mathrm{e}{5}}$  & ${8\mathrm{e}{5}}$  & ${3\mathrm{e}{6}}$ \\
${}^{1.0}\Ans$  & $\mathbf{0.50}$  & $\mathbf{0.38}$  & $\mathbf{0.35}$  & $\mathbf{0.36}$  & $\mathbf{0.40}$  & $\mathbf{0.49}$  & $\mathbf{0.65}$  & $\mathbf{0.90}$  & $\mathbf{1.3}$  & $\mathbf{2.1}$  & $\mathbf{3.3}$  & $\mathbf{5.6}$  & $\mathbf{9.9}$  & $\mathbf{18}$  & $\mathbf{34}$  & $\mathbf{65}$  & $\mathbf{130}$  & $\mathbf{266}$  & $\mathbf{559}$ \\
\hline\hline
${}^{0.99}\Zn$  & $\mathbf{0.49}$  & ${0.43}$  & ${0.43}$  & ${0.49}$  & ${0.63}$  & ${0.86}$  & ${1.3}$  & ${2.0}$  & ${3.2}$  & ${5.5}$  & ${10.}$  & ${19}$  & ${36}$  & ${73}$  & ${149}$  & ${322}$  & ${699}$  & ${2\mathrm{e}{3}}$  & ${4\mathrm{e}{3}}$  \\
${}^{0.99}\Ans$  & $\mathbf{0.49}$  & $\mathbf{0.35}$  & $\mathbf{0.29}$  & $\mathbf{0.28}$  & $\mathbf{0.30}$  & $\mathbf{0.35}$  & $\mathbf{0.44}$  & $\mathbf{0.59}$  & $\mathbf{0.83}$  & $\mathbf{1.3}$  & $\mathbf{2.0}$  & $\mathbf{3.2}$  & ${5.5}$  & ${9.7}$  & ${18}$  & ${33}$  & ${65}$  & ${131}$  & ${268}$  \\
${}^{0.99}\randn$  & ${2.3}$  & ${1.5}$  & ${1.1}$  & ${0.93}$  & ${0.87}$  & ${0.89}$  & ${0.97}$  & ${1.1}$  & ${1.4}$  & ${1.8}$  & ${2.4}$  & ${3.4}$  & $\mathbf{5.1}$  & $\mathbf{7.7}$  & $\mathbf{12}$  & $\mathbf{20}$  & $\mathbf{33}$  & $\mathbf{56}$  & $\mathbf{99}$  \\
\hline\hline
${}^{0.95}\Zn$  & $\mathbf{0.48}$  & ${0.36}$  & ${0.33}$  & ${0.35}$  & ${0.42}$  & ${0.54}$  & ${0.75}$  & ${1.1}$  & ${1.8}$  & ${2.9}$  & ${5.0}$  & ${8.9}$  & ${16}$  & ${32}$  & ${63}$  & ${130}$  & ${274}$  & ${590}$  & ${1\mathrm{e}{3}}$  \\
${}^{0.95}\Ans$  & $\mathbf{0.48}$  & $\mathbf{0.31}$  & $\mathbf{0.26}$  & $\mathbf{0.24}$  & $\mathbf{0.25}$  & $\mathbf{0.29}$  & $\mathbf{0.36}$  & $\mathbf{0.47}$  & $\mathbf{0.66}$  & $\mathbf{0.98}$  & $\mathbf{1.5}$  & ${2.5}$  & ${4.2}$  & ${7.3}$  & ${13}$  & ${25}$  & ${47}$  & ${95}$  & ${192}$  \\
${}^{0.95}\randn$  & ${1.5}$  & ${0.95}$  & ${0.72}$  & ${0.61}$  & ${0.57}$  & ${0.58}$  & ${0.63}$  & ${0.74}$  & ${0.91}$  & ${1.2}$  & ${1.6}$  & $\mathbf{2.2}$  & $\mathbf{3.3}$  & $\mathbf{5.0}$  & $\mathbf{7.9}$  & $\mathbf{13}$  & $\mathbf{21}$  & $\mathbf{36}$  & $\mathbf{64}$  \\
\hline\hline
${}^{0.90}\Zn$  & $\mathbf{0.45}$  & ${0.31}$  & ${0.28}$  & ${0.28}$  & ${0.33}$  & ${0.41}$  & ${0.55}$  & ${0.80}$  & ${1.2}$  & ${2.0}$  & ${3.3}$  & ${5.8}$  & ${10}$  & ${20}$  & ${38}$  & ${77}$  & ${160}$  & ${339}$  & ${748}$ \\
${}^{0.90}\Ans$  & $\mathbf{0.45}$  & $\mathbf{0.29}$  & $\mathbf{0.23}$  & $\mathbf{0.21}$  & $\mathbf{0.22}$  & $\mathbf{0.25}$  & $\mathbf{0.31}$  & $\mathbf{0.40}$  & $\mathbf{0.56}$  & $\mathbf{0.82}$  & ${1.3}$  & ${2.0}$  & ${3.4}$  & ${5.9}$  & ${11}$  & ${20}$  & ${38}$  & ${75}$  & ${150}$ \\
${}^{0.90}\randn$  & ${1.2}$  & ${0.73}$  & ${0.55}$  & ${0.47}$  & ${0.44}$  & ${0.45}$  & ${0.49}$  & ${0.57}$  & ${0.70}$  & ${0.90}$  & $\mathbf{1.2}$  & $\mathbf{1.7}$  & $\mathbf{2.5}$  & $\mathbf{3.8}$  & $\mathbf{6.0}$  & $\mathbf{9.8}$  & $\mathbf{16}$  & $\mathbf{28}$  & $\mathbf{49}$ \\
\end{tabular}